# MATHEMATICAL MATRIX THEORY OF FIELD IN N-DIMENSIONAL METRIC SPACE: RIGOROUS DERIVATION OF THE EQUATIONS OF THE FIELD WITH APPLICATION IN ELECTROMAGNETIC-GRAVITATIONAL FIELDS.

**Alexander D. Dymnikov**

Louisiana Accelerator Center/Physics Dept/, Louisiana University at Lafayette, 320 Cajundome Blvd., Lafayette Louisiana 70506, United States
Tel: 337-482-1845
Fax: 337-482-6190
E-mail: dymnikov@louisiana.edu

## ABSTRACT

During the last century the tensor theory of the gravitational field was developed. We propose and develop the novel, pure mathematical, matrix theory of the field in $n$-dimensional metric space. The definition of the mathematical $n \times n$ field matrix and the equations of motion of the mathematical point are given. The interpretation of the nature of the mathematical field and the mathematical points can be different and depends on our knowledge of the nature. It is shown that the equations of motion are different for symmetric and antisymmetric field matrices. In the matrix field theory the equations of the field are rigorously derived. This theory reveals that in the 4-dimensional metric space the field matrix is the electromagnetic-gravitational field matrix, where the antisymmetric part is the matrix of electromagnetic field and the symmetric part is the gravitational field matrix. The partial cases of this matrix are electric-gravitational, magnetic-gravitational and gravitational field matrices. It is shown that the elements of all obtained matrices are the Christoffel symbols of the first and the second order or their derivatives.

In the matrix theory we find the eigenvalues of the matrix $g^{-1}R$, where $g$ is the metric matrix and $R$ is the Ricci matrix. There are two different types of the field matrix equations. In the first type all eigenvalues of the matrix $g^{-1}R$ are equal and the appropriate eigenvector is equal to the velocity vector. When we compare every two different eigenvalues the differential equations appears. It is shown that in the first type the metric matrix is determined by solving these differential equations. Solving these equations we obtain some constants of integration which can be interpreted as moving mass, concentrated mass or density. In the second type for only one eigenvalue the eigenvector is equal to the velocity vector. Both types of the field matrix equations resemble the Maxwell equations for the electromagnetic field. For the electromagnetic field we obtain exact Maxwell equations.

The absolute (invariant) velocities and field matrix in 4-dimensional space become non-integrable but differentiable if we have nonzero density of mass or charge and (or) we have concentrated mass or charge.

The use of the first type field matrix equations gave us the possibility to find the general spherically symmetric solution of the metric matrix equation for the gravitational field. This new solution describes the motion of moving mass in the field of concentrated mass and density. The weak field solution is obtained as a partial case of the general spherically symmetric solution when the moving mass is zero. It is shown that the Schwarzschild weak solution is not exactly correct.

The obtained second type of the matrix field equations was solved for the Friedmann-Lobachevsky model of the metric matrix $g$. We obtained the expression for the density of particles. The comparison of this density expression with the cosmic microwave background spectrum from COBE shows that this expression can be used as the mathematical model of the Big Bang.

## INTRODUCTION

The Nobel Prize winner E. Wigner writes in his article "The Unreasonable Effectiveness of Mathematics in the Natural Sciences": "The first point is that the enormous usefulness of mathematics in the natural sciences is something bordering on the mysterious and there is no rational explanation for it. Second, it is just this uncanny usefulness of mathematical concepts that raises the question of the uniqueness of our physical theories."

In this paper we develop the matrix theory of *n*-dimensional mathematical field and the motion of mathematical points in this field. This theory is a pure mathematical theory and for every application we need to interpret the obtained results.

Two spaces are considered: the *n*-dimensional space of an integrable coordinate vector $\{x\}$ with the metric $g(x)$ and the *n*-dimensional space of an non-integrable but differentiable coordinate vector $\{Q\}$ with the metric $g(Q)$, which is unit matrix *I*.

$$dQ = e\,dx,$$

$$d\tau \equiv \sqrt{d\tilde{x}\,g(x)\,dx} = \sqrt{d\tilde{Q}\,g(Q)\,dQ},$$

$$g(x) = \tilde{e}\,e,\quad g(Q) = I.$$

Here *e* is a non-integrable but differentiable frame matrix. The derivatives of the matrix function *e* are expressed through the Christoffel matrices, which elements are the Christoffel symbols of the first and the second kind, and through the Ricci and Riemann curvature matrices, which elements are the elements of Ricci and Riemann curvature tensors. We call the coordinate space of the non-integrable but differentiable coordinate vector {Q}, as the *absolute* (invariant) space. The *absolute* (invariant) velocity vector

$$u(Q) \equiv dQ/d\tau$$

and the *absolute* (invariant) field matrix

$$P \equiv u(Q)\,\tilde{\partial}(Q)$$

are introduced. We obtain two groups of the matrix field equations, the first of which is written in the two following forms:

$$\partial(Q)\langle P\rangle - \tilde{P}\,\partial(Q) = \rho\,u(Q),$$

and

$$K\,u(Q) = \rho\,u(Q),$$

where $\langle P\rangle$ is the trace of the matrix *P*,

$$K \equiv ([D\,\tilde{D}]e)\,e^{-1},$$

$$[D\,\tilde{D}]_{ij} \equiv D_i D_j - D_j D_i, \quad i,j = 1,2,...,n.$$

$$D_k \equiv \partial_k(Q) = \frac{\partial}{\partial Q_k}$$

Here $K$ is the *absolute* Ricci matrix function, $u(Q)$ is the $n$-dimensional *absolute* (invariant) velocity vector, $\rho$ is a scalar function, which is the eigenvalue of $K$ with the corresponding eigenvector $u(Q)$. Applications to the four-dimensional Riemannian space-time are considered. It is shown that in this case the *absolute* (invariant) field matrix can be considered as the *absolute* electromagnetic-gravitational field matrix. The symmetric part of this matrix can be considered as the *absolute* gravitational field matrix. The antisymmetric part can be considered as the *absolute* electromagnetic field matrix. The differential equations for these two matrices and for the metric matrix were obtained. The partial case of these equations is the Maxwell equations for the electromagnetic field.

Six general solutions for the static spherically symmetric gravitational field are found. The first solution is the most general and it contains three constants, which can be interpreted as two masses and density. It describes the motion of the mass $m$ in the field of the concentrated mass $M$ and the density $\rho$.

The obtained matrix field equations were solved also for the Friedmann-Lobachevsky model of the metric $g$. We obtained the expression for the density of particles which looks like the expression describing the mathematical model of the Big Bang, Our expression is compared with the experimental cosmic microwave background spectrum from COBE.

# 1. VECTOR AND MATRIX FUNCTIONS AND DIFFERENTIAL OPERATORS.

## 1.1 Matrix notations.

The square $n \times n$ matrix as a whole is denoted by $a$, where

$$a \equiv \begin{pmatrix} a_{11} & a_{12} & ... & a_{1n} \\ a_{21} & a_{22} & ... & a_{2n} \\ ... & ... & ... & ... \\ a_{n1} & a_{n2} & ... & a_{nn} \end{pmatrix}.$$

We will write the matrix $a$ also in a horizontal alignment (notations from Mathematica by S. Wolfram):

$$a \equiv \{\{a_{11}, a_{12}, ..., a_{1n}\}, ..., \{a_{n1}, a_{n2}, ..., a_{nn}\}\}.$$

The inverse matrix of $a$ is denoted by $a^{-1}$. Transpose matrix:

$$\tilde{a} \equiv \{\{a_{11}, a_{21}, ..., a_{n1}\}, ..., \{a_{1n}, a_{2n}, ..., a_{nn}\}\} \equiv \begin{pmatrix} a_{11} & a_{21} & ... & a_{n1} \\ a_{12} & a_{22} & ... & a_{n2} \\ ... & ... & ... & ... \\ a_{1n} & a_{2n} & ... & a_{nn} \end{pmatrix}$$

The trace of a matrix $a$ is denoted by $\langle a \rangle$,

$$\langle a \rangle \equiv a_{\lambda\lambda} = a_{11} + a_{22} + ... + a_{nn}.$$

**1.2 Coordinate vectors $x$ and $Q$.**

The mathematical point is described by the set of *n* parameters (or coordinates). The position (the set of *n* parameters) of any mathematical point is described by an *n*-coordinate vectors $\{x\}$ or $\{Q\}$ where

$$\{x\} = \{x_1, x_2, ..., x_n\},$$

$$\{Q\} = \{Q_1, Q_2, ..., Q_n\},$$

Column vectors are written as

$$x \equiv \{\{x_1\}, \{x_2\}, ..., \{x_n\}\} \equiv \begin{pmatrix} x_1 \\ x_2 \\ ... \\ x_n \end{pmatrix},$$

$$Q \equiv \{\{Q_1\}, \{Q_2\}, ..., \{Q_n\}\} \equiv \begin{pmatrix} Q_1 \\ Q_2 \\ ... \\ Q_n \end{pmatrix},$$

and row vectors as

$$\tilde{x} \equiv \{\{x_1, x_2, ...x_n\}\} \equiv (x_1, x_2, ..., x_n),$$

$$\tilde{Q} \equiv \{\{Q_1, Q_2, ...Q_n\}\} \equiv (Q_1, Q_2, ..., Q_n)$$

**1.3 Differential vector- operators $\partial(x)$ and $\partial(Q)$.**

For the *n*-vectors $\{x\}$ and $\{Q\}$ the differential operator *n*-vectors $\{\partial(x)\}$ and $\{\partial(Q)\}$

$$\{\partial(x) \equiv (\partial_1(x), ..., \partial_n(x)),$$
$$\{\partial(Q) \equiv (\partial_1(Q), ..., \partial_n(Q)),$$

can be written as the column vectors $\partial(x)$ and $\partial(Q)$

$$\partial(x) \equiv \{\{\partial_1(x)\}, ..., \{\partial_n(x)\}\} = \begin{pmatrix} \partial_1(x) \\ ... \\ \partial_n(x) \end{pmatrix},$$

$$\partial(Q) \equiv \{\{\partial_1(Q)\}, ..., \{\partial_n(Q)\}\} = \begin{pmatrix} \partial_1(Q) \\ ... \\ \partial_n(Q) \end{pmatrix},$$

or as the row vectors $\tilde{\partial}(x)$ and $\tilde{\partial}(Q)$,

$$\tilde{\partial}(x) \equiv \{\{\partial_1(x), ..., \partial_n(x)\}\} = (\partial_1(x),...,\partial_n(x)),$$

$$\tilde{\partial}(Q) \equiv \{\{\partial_1(Q), ..., \partial_n(Q)\}\} = (\partial_1(Q),...,\partial_n(Q)),$$

where

$$\partial_i(x) \equiv \frac{\partial}{\partial x_i},$$

$$\partial_i(Q) \equiv \frac{\partial}{\partial Q_i}.$$

### 1.4 Scalar differential operator and exact differential.

The scalar differential operator $d$ is defined by the following way:

$$d = \tilde{\partial}(x)\, dx = \tilde{\partial}(Q)\, dQ.$$

The exact differentials *dQ* and *dx* are defined by the expressions

$$d\,Q = Q\,d = Q\,\tilde{\partial}(x)\,dx = e\,d\,x, \quad e \equiv Q\,\tilde{\partial}(x),$$

$$d\,x = x\,d = x\,\tilde{\partial}(Q)\,dQ.$$

The expression $Q\,\tilde{\partial}(x)$, where $Q$ is an *n*-vector function of *x*,

$$Q\,\tilde{\partial}(x) \equiv \{\{\partial_1(x)Q_1,\dots,\partial_n(x)Q_1\},\dots,\{\partial_1(x)Q_n,\dots,\partial_n(x)Q_n\}\},$$

stands for the matrix whose *ik*-th element is equal to $\partial Q_i/\partial x_k$. The expression $x\,\partial(Q)$ denotes the matrix

$$x\,\tilde{\partial}(Q) \equiv \{\{\partial_1(Q)\,x_1,\dots,\partial_n(Q)\,x_1\},\dots,\{\partial_1(Q)\,x_n,\dots,\partial_n(Q)\,x_n\}\},$$

whose *ik*-th element is equal to $\partial x_i/\partial Q_k$. The frame matrix *e* is the matrix

$$e \equiv Q\,\vec{\tilde{\partial}}(x)$$

with the elements

$$e_{ik} = \partial Q_i / \partial x_k.$$

From the expressions for *dQ* and *dx* it follows

$$dQ = e\,dx = e\,x\,\tilde{\partial}(Q)\,dQ,$$

$$x\tilde{\partial}(Q) = e^{-1}.$$

## 1.5 The contravariant and covariant vectors.

From the expression

$$d = \tilde{\partial}(Q)\,dQ = \tilde{\partial}(Q)\,e\,dx = \tilde{\partial}(x)\,dx = \tilde{\partial}(x)\,e^{-1}\,dQ,$$

we will find the differential $n$-vector operators $\partial(Q)$ and $\partial(x)$,

$$.\partial(Q) = \tilde{e}^{-1}\partial(x), \quad \partial(x) = \tilde{e}\,\partial(Q).$$

The vectors $dQ$ and $dx$,

$$dQ = e\,dx, \quad dx = e^{-1}dQ,$$

are the contravariant vectors and the vectors $\partial(x)$ and $\partial(Q)$ are the covariant vectors.
The relation

$$dQ = e\,dx,$$

links the contravariant coordinate vector $dQ$ and the contravariant coordinate vector $dx$.
The relation

$$\partial(Q) = \tilde{e}^{-1}\partial(x)$$

connects the covariant coordinate vector operator $\partial(Q)$ and the covariant vector operator $\partial(x)$.

Vectors $dQ$ and $dx$ are contravariant vectors and vector - operators $\partial(Q)$ and $\partial(x)$ are covariant vectors.

### 1.6 Four coordinate forms of any absolute matrix P.

We will use four coordinate forms of any absolute matrix $P$: the twice contravariant matrix $\overline{P}$, the twice covariant matrix $\underline{P}$, the covariant-contravariant matrix $\underset{\rightarrow}{P}$ and the contravariant-covariant matrix $\vec{P}$, where

$$\overline{P} \equiv e^{-1} P \tilde{e}^{-1},$$

$$\underline{P} \equiv \tilde{e} P e,$$

$$\underset{\rightarrow}{P} \equiv \tilde{e} P \tilde{e}^{-1} = g\,\overline{P} = \underline{P}\,g^{-1}.$$

$$\vec{P} \equiv e^{-1} P e = \overline{P}\,g = g^{-1}\,\underline{P}.$$

### 1.7 The contravariant and covariant indices.

In the matrix theory the indices are usually not shown. If it is necessary the following notations for matrix or vector indices are used: $\underline{k}$ – covariant index, $\overline{k}$ - contravariant index. Some vectors and matrices with indices are shown below.

$$\partial_m(x) = \partial_{\underline{m}}(x),$$

$$x_m = x_{\overline{m}},$$

$$(Q\partial(x))_{mn} = (Q\partial(x))_{\overline{m}\underline{n}} = \frac{\partial Q_{\overline{m}}}{\partial x_{\underline{n}}}.$$

$$\overline{P}_{mn} = \overline{P}_{\overline{m}\overline{n}}, \quad \underline{P} = \underline{P}_{\underline{m}\underline{n}}, \quad \underset{\rightarrow}{P}_{mn} = \underset{\rightarrow}{P}_{\underline{m}\overline{n}}, \quad \vec{P}_{mn} = \vec{P}_{\overline{m}\underline{n}}$$

The matrix $e = Q\tilde{\partial}(x)$ is a contravariant-covariant matrix.

### 1.8 Unit vectors and matrices.

We introduce the contravariant unit vector

$$\bar{i}(\lambda) \equiv \partial_\lambda x,$$

covariant unit vector

$$\underline{i}(\lambda) \equiv \partial x_\lambda,$$

and the unit vector $i(\lambda)$, where

$$i_j(\lambda) = \delta_{j\lambda}.$$

If we are not interested in the nature of the unit vector, we use the unit vector $i(\lambda)$. As an example, for $n = 4$

$$i(1) = \{1,0,0,0\}, \quad i(2) = \{0,1,0,0\}, \quad i(3) = \{0,0,1,0\}, \quad i(4) = \{0,0,0,1\}.$$

From the contravariant and covariant unit vectors it is easy to obtain four different unit matrices,

$$\bar{i}(\lambda)\tilde{\bar{i}}(\lambda), \quad \bar{i}(\lambda)\tilde{\underline{i}}(\lambda), \quad \underline{i}(\lambda)\tilde{\bar{i}}(\lambda), \quad \underline{i}(\lambda)\tilde{\underline{i}}(\lambda).$$

If we are not interested in the nature of the unit matrix, we use the following definition of the unit (or identity) matrix:

$$I_n = i(\mu)\,\tilde{i}(\mu) = i(1)\,\tilde{i}(1) + i(2)\,\tilde{i}(2) + ... + i(n)\,\tilde{i}(n).$$

As an example, for $n = 4$,

$$I_4 = i(1)\,\tilde{i}(1) + i(2)\,\tilde{i}(2) + i(3)\,\tilde{i}(3) + i(4)\,\tilde{i}(4) = \begin{pmatrix} 1 & 0 & 0 & 0 \\ 0 & 1 & 0 & 0 \\ 0 & 0 & 1 & 0 \\ 0 & 0 & 0 & 1 \end{pmatrix}.$$

### 1.9 Short operator notations.

In many cases we will use the short notations,

$$\partial \equiv \{\{\partial_1\},...,\{\partial_n\}\}, \quad \tilde{\partial} \equiv \{\{\partial_1,...,\partial_n\}\}, \quad \partial_k \equiv \partial_k(x) = \frac{\partial}{\partial x_k}.$$

$$D \equiv \{\{D_1\},...,\{D_n\}\}, \quad \tilde{D} \equiv \{\{D_1,...,D_n\}\}, \quad D_k \equiv \partial_k(Q) = \frac{\partial}{\partial Q_k}$$

$$Q\,\tilde{\partial} \equiv \{\{\partial_1 Q_1,...,\partial_n Q_1\},...,\{\partial_1 Q_n,...,\partial_n Q_n\}\}, \quad Q_i \partial_k \equiv \partial Q_i/\partial x_k,$$

$$\partial\tilde{Q} \equiv \{\{\partial_1 Q_1,...,\partial_1 Q_n\},...,\{\partial_n Q_1,...,\partial_n Q_n\}\}, \quad \partial_i Q_k \equiv \partial Q_k/\partial x_i.$$

The scalar differential operator is $d$, where

$$d \equiv \tilde{\partial} dx = d\tilde{x}\partial = \frac{\partial}{\partial x_\alpha} dx_\alpha,$$

$$d \equiv \tilde{D}\, dQ = d\tilde{Q}\, D = \frac{\partial}{\partial Q_\alpha} dQ_\alpha.$$

The summation is understood to be carried out over any *repeated* Greek index symbol that appears twice in any expression.

### 1.10 The compatibility conditions.

In order that the differential $dQ$ be exact differentials, it is necessary and sufficient that there are the relations:

$$\partial_k(x)\, \partial_j(x)\, Q - \partial_j(x)\, \partial_k(x)\, Q = \partial_k\, e\, i(j) - \partial_j\, e\, i(k) = 0, \quad k, j = 1,...,n$$

where

$$e \equiv Q\vec{\tilde{\partial}}(x),$$

with the elements

$$e_{ik} = \partial Q_i / \partial x_k.$$

The equations

$$\partial_k\, e\, i(j) - \partial_j\, e\, i(k) = 0, \quad k, j = 1,...,n$$

are called the compatibility conditions.

## 1.11 The antisymmetric matrix-differential operators.

The antisymmetric matrix - differential operators $[\partial \tilde{\partial}]$ and $[D\tilde{D}]$,

$$[\partial \tilde{\partial}] \equiv \{\{[\partial \tilde{\partial}]_{11},[\partial \tilde{\partial}]_{12},...,[\partial \tilde{\partial}]_{1n}\},...,\{[\partial \tilde{\partial}]_{n1},[\partial \tilde{\partial}]_{n2},...,[\partial \tilde{\partial}]_{nn}\}\},$$

$$[D\tilde{D}] \equiv \{\{[D\tilde{D}]_{11},[D\tilde{D}]_{12},...,[D\tilde{D}]_{1n}\},...,\{[D\tilde{D}]_{n1},[D\tilde{D}]_{n2},...,[D\tilde{D}]_{nn}\}\},$$

which are called the compatibility matrix - operators, are defined by the elements

$$[\partial \tilde{\partial}]_{ij} \equiv \partial_i \partial_j - \partial_j \partial_i, \quad i,j=1,2,...,n.$$

$$[D\tilde{D}]_{ij} \equiv D_i D_j - D_j D_i, \quad i,j=1,2,...,n$$

The compatibility matrix - operators may be written in the form

$$[\partial \tilde{\partial}] = [\partial \tilde{\partial}]_{\alpha \beta} \, i(\alpha) \, \tilde{i}(\beta).$$

$$[D\tilde{D}] = [D\tilde{D}]_{\alpha \beta} \, i(\alpha) \, \tilde{i}(\beta).$$

## 1.12 Two forms of the compatibility conditions.

The vector function $Q$ has the full (or exact) differential $dQ$,

$$dQ = e\,dx, \quad e = Q\tilde{\partial}, \quad Q\tilde{\partial}\, i(j) = \partial_j Q = e\,i(j)$$

which satisfies the compatibility conditions for the vector $Q$,

$$[\partial\tilde{\partial}]_{kj} Q = (\partial_k \partial_j - \partial_j \partial_k) Q = \partial_k e\,i(j) - \partial_j e\,i(k) = 0, \quad k, j = 1,\ldots,n.$$

The vector function $x$ has the full (or exact) differential $dx$,

$$dx = e^{-1} dQ, \quad e^{-1} = x\tilde{D}, \quad x\tilde{D}i(j) = e^{-1}i(j) = D_j x$$

which satisfies the compatibility conditions for the vector $x$

$$[D\tilde{D}]_{kj} x = (D_k D_j - D_j D_k) x = D_k e^{-1} i(j) - D_j e^{-1} i(k) = 0, \quad k, j = 1,\ldots,n.$$

Therefore there are two forms of the compatibility conditions:

$$[\partial\tilde{\partial}]_{kj} Q = (\partial_k \partial_j - \partial_j \partial_k) Q = \partial_k e\,i(j) - \partial_j e\,i(k) = 0, \quad k, j = 1,\ldots,n$$

and

$$[D\tilde{D}]_{kj} x = (D_k D_j - D_j D_k) x = D_k e^{-1} i(j) - D_j e^{-1} i(k) = 0, \quad k, j = 1,\ldots,n.$$

For the operator $[D\tilde{D}]_{m\mu}$, using the compatibility conditions,

$$\partial_k e\,i(j) - \partial_j e\,i(k) = 0, \quad k, j = 1,\ldots,n$$

and the connection between the covariant coordinate operator vector $\partial \equiv \partial(x)$ and the *absolute* operator vector $D$,

$$D = \tilde{e}^{-1} \partial, \quad D_m = \tilde{i}(m)\tilde{e}^{-1}\partial = \tilde{e}^{-1}_{m\lambda} \partial_\lambda,$$

one obtains

$$[D\tilde{D}]_{m\mu} = D_m D_\mu - D_\mu D_m =$$
$$(D_m \tilde{e}^{-1}_{\mu\lambda})\partial_\lambda - (D_\mu \tilde{e}^{-1}_{m\lambda})\partial_\lambda + \tilde{e}^{-1}_{m\alpha}\tilde{e}^{-1}_{\mu\beta}[\partial\tilde{\partial}]_{\alpha\beta} = \tilde{e}^{-1}_{m\alpha}\tilde{e}^{-1}_{\mu\beta}[\partial\tilde{\partial}]_{\alpha\beta},$$

or, in the matrix form

$$[D\tilde{D}] = i(\alpha)\tilde{i}(\beta)([D\tilde{D}]_{\alpha\beta} = \tilde{e}^{-1}[\partial\tilde{\partial}]\,e^{-1} = \tilde{e}^{-1}i(\alpha)\tilde{i}(\beta)([\partial\tilde{\partial}]_{\alpha\beta}\,e^{-1})\,.$$

The last expression is the transformation from the twice covariant coordinate differential operator matrix $[\partial\tilde{\partial}]$ to the *absolute* twice covariant differential operator matrix $[D\tilde{D}]$.

The expression

$$[\partial\tilde{\partial}] = \tilde{e}\,[D\tilde{D}]\,e$$

is the transformation from the *absolute* twice covariant differential operator matrix $[D\tilde{D}]$ to the twice covariant coordinate differential operator matrix $[\partial\tilde{\partial}]$.

In general case we consider the vector *x* as an integrable vector and the vector *Q* as an nonintegrable vector.

**1.13 The absolute coordinate vector, the absolute frame matrix and the absolute coordinate space**

The connection between the exact differentials *dQ* and *dx* is defined by the expressions:

$$dQ = Q\tilde{\partial}(x)\,dx = e\,dx, \qquad e = Q\tilde{\partial}(x),$$

$$dx = x\tilde{\partial}(Q)\,dQ = e^{-1}\,dQ, \qquad e^{-1} = x\tilde{\partial}(Q),$$

$$d\tilde{Q}\,dQ = d\tilde{x}\,\tilde{e}\,e\,dx,$$

$$\tilde{e}\,e = g(x),$$

where $e$ is –a non-integrable but differentiable frame matrix, $g(x)$ is the integrable metric matrix of the coordinate space $x$. We call the coordinate vector $Q$ as the *absolute* coordinate vector and the frame matrix $e$ as the *absolute* frame matrix.

The non-integrable but differentiable *absolute* frame matrix $e$ can be written in the form

$$e = \Omega\,e_0,$$

where $e_0$ is an integrable frame matrix and $\Omega$ is a non-integrable but differentiable orthogonal matrix,

$$\tilde{\Omega}\,\Omega = I_n.$$

Any $n \times n$ matrix function $F$ is integrable, if

$$(\partial_\alpha \partial_\beta - \partial_\beta \partial_\alpha)F = 0, \quad \alpha, \beta = 1,...,n.$$

Any $n \times n$ matrix function $F$ is not integrable, if

$$(\partial_\alpha \partial_\beta - \partial_\beta \partial_\alpha)F \neq 0, \quad \alpha, \beta = 1,...,n.$$

The integrability of the matrix $e_0$

$$e_0 = \tilde{\Omega}\,e$$

means that

$$(\partial_\alpha \partial_\beta - \partial_\beta \partial_\alpha) e_0 = 0.$$

From the expression for the infinitesimal invariant interval,

$$d\tau^2 = d\tilde{x}\, g(x)\, dx = d\tilde{x}\, \tilde{e}\, e\, dx = d\tilde{x}\, \tilde{e}_0\, e_0\, dx,$$

it follows that

$$d\tau^2 = d\tilde{Q}\, dQ = d\tilde{Q}\, g(Q)\, dQ, \quad g(Q) = I_n,$$

where the metric matrix $g(Q)$ is the unit matrix $I_n$. We call the space of the non-integrable but differentiable *absolute* coordinate vector $Q$ as the *absolute* coordinate space.

## 2. DERIVATIVES OF THE NON-INTEGRABLE BUT DIFFERENTIABLE FRAME MATRIX.

### 2.1 One-index matrices.

The *absolute* non-integrable $n \times n$ frame matrix $e$ satisfies the compatibility conditions,

$$\boxed{\partial_j\, e\, i(k) = \partial_k\, e\, i(j), \quad j, k = 1,....,n},$$

where the matrices

$$\partial_j e, \quad j = 1,...,n$$

are one-index matrices and where

$$\tilde{e}\, e = \tilde{e}_0\, e_0 = g(x),$$

$$e = \Omega e_0, \quad \tilde{\Omega}\Omega = I_n$$

Let us consider the one-index matrix function $h^c$, where

$$h^c \equiv \tilde{e}\, \partial_c e, \quad c = 1, 2,...,n, \quad h^c_{ab} = \tilde{i}(a)\, \tilde{e}\, \partial_c e\, i(b),$$

and $e$ is the matrix function. From the compatibility conditions

$$\partial_j e\, i(k) = \partial_k e\, i(j), \quad j, k = 1,....,n,$$

we obtain the following relations between the elements of the matrix $h^c$:

$$h^c_{ab} = \tilde{i}(a)\, \tilde{e}\, \partial_c e\, i(b) = \tilde{i}(a)\, \tilde{e}\, \partial_b e\, i(c) = h^b_{ac},$$

$$h^a_{bc} = \tilde{i}(b)\, \tilde{e}\, \partial_a e\, i(c) = \tilde{i}(b)\, \tilde{e}\, \partial_c e\, i(a) = h^c_{ba}$$

$$h^a_{cb} = \tilde{i}(c)\, \tilde{e}\, \partial_a e\, i(b) = \tilde{i}(c)\, \tilde{e}\, \partial_b e\, i(a) = h^b_{ca}$$

It is possible to present the matrix function $h^c$ as the sum of two matrix functions with different symmetry

$$h^c = \gamma^c + \lambda^c,$$

with

$$\lambda^c_{ab} \equiv \frac{1}{2}(h^c_{ab} - h^b_{ac} + h^a_{cb} - h^b_{ca} + h^a_{bc} - h^c_{ba}),$$

and

$$\gamma^c_{ab} \equiv \frac{1}{2}(h^c_{ab} + h^b_{ac} + h^b_{ca} - h^a_{bc} - h^a_{cb} + h^c_{ba}).$$

Here $\gamma^c_{ab}$ has symmetry in $c$ and $b$ indices,

$$\boxed{\gamma^c_{ab} = \gamma^b_{ac}},$$

and matrix $\lambda^c$ is an antisymmetric matrix,

$$\boxed{\lambda^c = -\tilde{\lambda}^c, \quad \lambda^c_{ab} = -\lambda^c_{ba}}.$$

where the element $\gamma^c_{ab}$ has symmetry in $c$ and $b$ indices.

From the equations

$$h^a_{bc} = h^c_{ba}, \quad h^c_{ab} = h^b_{ac}, \quad h^a_{cb} = h^b_{ca}$$

it follows, that the matrix $\lambda^c$ is zero matrix,

$$\lambda^c = 0.$$

Therefore

$$h^c = \gamma^c,$$

$$h^c = \gamma^c, \quad \gamma^c_{ab} \equiv \frac{1}{2}(h^c_{ab} + h^b_{ac} + h^b_{ca} - h^a_{bc} - h^a_{cb} + h^c_{ba}),$$

or

$$(\tilde{e}\,\partial_c e)_{ab} = \gamma^c_{ab} = \frac{1}{2}((h^c_{ab} + h^c_{ba}) + (h^b_{ca} + h^b_{ac}) - (h^a_{bc} + h^a_{cb})) \quad c = 1, 2, ..., n,$$

From the last equation and from the equations,

$$\tilde{e}\,e = g(x)$$

$$\partial_c g_{ab} = h^c_{ab} + h^c_{ba}, \quad \partial_b g_{ac} = h^b_{ac} + h^b_{ca}, \quad \partial_a g_{bc} = h^a_{bc} + h^a_{cb},$$

it follows,

$$\tilde{e}\,\partial_c e = \gamma^c = \frac{1}{2}(\partial_c g + g\,i(c)\tilde{\partial} - \partial\,\tilde{i}(c)\,g).$$

### 2.2 The one-index Christoffel matrices of the first kind and of the second kind.

In the above equations the matrix elements $\gamma^c_{ab}$ are the *Christoffel symbols* $\Gamma_{a,cb}$ of *the first kind*, which are used in the general theory of relativity,

$$\gamma_{ab}^c = \tilde{i}(a)\gamma^c i(b) = \frac{1}{2}(\partial_c g_{ab} + \partial_b g_{ac} - \partial_a g_{bc}) = \Gamma_{a,cb}.$$

We call this matrix $\gamma^c$ as the *one-index Christoffel matrix of the first kind*. We call the matrix $\sigma^m$, where

$$\sigma^m \equiv e^{-1}\partial_m e = g^{-1}\gamma^m$$

$$\partial_m e = \tilde{e}^{-1}\gamma^m = e\sigma^m$$

$$\tilde{e}\partial_m e = \gamma^m = g\sigma^m$$

as the one-index *Christoffel matrix of the second kind*. The matrices $\gamma^m$ and $\sigma^m$ have the following property,

$$\gamma^m i(k) = \gamma^k i(m),$$

$$\sigma^m i(k) = \sigma^k i(m).$$

### 2.3 The second derivatives of the non-integrable but differentiable matrices *e*.

Using the first derivatives it is easy to obtain the second derivatives of the non-integrable but differentiable matrices *e* and $e^{-1}$:

$$\partial_a \partial_b e = e(\partial_a \sigma^b + \sigma^a \sigma^b), \quad a = 1,...,n-1; \quad b = 2,...,n; \quad b > a$$

$$\partial_a \partial_b e^{-1} = (-\partial_a \sigma^b + \sigma^b \sigma^a)e^{-1}$$

$$a = 1,...,n-1; \quad b = 2,...,n; \quad b > a$$

Acting by the differential compatibility operator

$$[\partial \tilde{\partial}]_{ab} = \partial_a \partial_b - \partial_b \partial_a$$

on the frame matrix $e$, we obtain

$$[\partial \tilde{\partial}]_{ab} e = e \sigma^{ab},$$

where the two-index matrix $\sigma^{ab}$ is defined by the expression,

$$\sigma^{ab} \equiv \partial_a \sigma^b - \partial_b \sigma^a + \sigma^a \sigma^b - \sigma^b \sigma^a = e^{-1} [\partial \tilde{\partial}]_{ab} e,$$

$$a = 1,...,n-1; \quad b = 2,...,n; \quad b > a$$

$$\sigma^{ab}_{mk} = \sigma^{\underline{ab}}_{m\underline{k}}$$

The Riemann curvature tensor is identified as

$$R^m_{kab} = \partial_a \Gamma^m_{bk} - \partial_b \Gamma^m_{ak} + \Gamma^m_{as} \Gamma^s_{bk} - \Gamma^m_{bs} \Gamma^s_{ak}$$

It is easy to see that the elements $\sigma^{ab}_{mk}$ of the matrix $\sigma^{ab}$ are the elements of the Riemann curvature tensor $R^m_{kab}$,

$$\sigma^{ab}_{mk} = \partial_a \sigma^b_{mk} - \partial_b \sigma^a_{mk} + \sigma^a_{m\lambda} \sigma^b_{\lambda k} - \sigma^b_{m\lambda} \sigma^a_{\lambda k} =$$
$$\partial_a \Gamma^m_{bk} - \partial_b \Gamma^m_{ak} + \Gamma^m_{a\lambda} \Gamma^\lambda_{bk} - \Gamma^m_{b\lambda} \Gamma^\lambda_{ak} = R^m_{kab},$$

or in the matrix form,

$$\sigma^{ab} = g^{-1}i(\mu)\tilde{i}(\lambda)R_{\mu\lambda ab} = i(\mu)\tilde{i}(\lambda)R^{\mu}_{\lambda ab},$$

where

$$\sigma^m = i(\alpha)\tilde{i}(\beta)\Gamma^{\alpha}_{m\beta}.$$

We introduce also a two-index matrix $\gamma^{ab}$, where

$$\gamma^{ab} = \tilde{e}[\partial\tilde{\partial}]_{ab}e = g\,\sigma^{ab},$$

$$\gamma^{ab}_{mk} = \gamma^{\underline{ab}}_{\underline{mk}},$$

$$\gamma^{ab}_{mk} = g_{m\lambda}\,\sigma^{ab}_{\lambda k} = g_{m\lambda}(\partial_a\Gamma^{\lambda}_{bk} - \partial_b\Gamma^{\lambda}_{ak} + \Gamma^{\lambda}_{a\rho}\Gamma^{\rho}_{bk} - \Gamma^{\lambda}_{b\rho}\Gamma^{\rho}_{ak}) = g_{m\lambda}R^{\lambda}_{kab} = R_{mkab}.$$

In the last expression $R_{mkab}$ is the Riemann tensor with all lower indices.

Acting by the compatibility operator on the matrix $e$ and taking into a consideration that $e_0$ is the integrable function, we obtain

$$[\partial\tilde{\partial}]_{ab}e = \tilde{e}^{-1`}\gamma^{ab} = e\sigma^{ab},$$

$$[\partial\tilde{\partial}]_{ab}e = ([\partial\tilde{\partial}]_{ab}\Omega)e_0,$$

$$\Omega e_0\sigma^{ab} = ([\partial\tilde{\partial}]_{ab}\Omega)e_0.$$

The Riemann space-time becomes *flat* if the two-index matrix $\gamma^{ab}$ (or the two-index matrix $\sigma^{ab}$) becomes zero matrix.

From the two last equations it follows

$$e\sigma^{ab}e^{-1} = ([\partial\tilde{\partial}]_{ab}\Omega)\tilde{\Omega},$$

$$e_0\sigma^{ab}e_0^{-1} = \tilde{\Omega}([\partial\tilde{\partial}]_{ab}\Omega).$$

Therefore, the Riemannian spacetime becomes *flat* if and only if the orthogonal matrix $\Omega$ is *integrable*.

The integrability of the matrix function $e_0$ in the expression

$$\boxed{e = \Omega e_0},$$

implies

$$[\partial\tilde{\partial}]_{ab}\Omega = \Omega e_0 \sigma^{ab} e_0^{-1}.$$

From the last equation it follows that the orthogonal matrix $\Omega$ (or $e$) is integrable if and only if the two-index matrix $\sigma^{ab}$ (or the two-index matrix $\gamma^{ab}$) is a zero matrix. The integrability of the orthogonal matrix $\Omega$ leads to the integrability of the frame matrix $e$ and the vector $Q$.

## 3. THE ABSOLUTE RICCI, RIEMANN CURVATURE AND THE ABSOLUTE FORM OF THE BIANCHI IDENTITIES.

### 3.1 The *absolute* compatibility operator matrix.

The *absolute* compatibility two-index matrix operator $[D\tilde{D}]$ and the *coordinate* compatibility two-index matrix operator $[\partial\tilde{\partial}]_{ab}$ are defined by the elements

$$[D\tilde{D}]_{m\mu} \equiv D_m D_\mu - D_\mu D_m.$$

and

$$\left[\partial \tilde{\partial}\right]_{ab} = \partial_a \partial_b - \partial_b \partial_a .$$

These two operators are connected by the expression

$$[D\tilde{D}] = \tilde{e}^{-1}[\partial \tilde{\partial}]e^{-1} = \tilde{e}^{-1}(i(\alpha)\tilde{i}(\beta)[\partial \tilde{\partial}]_{\alpha\beta})e^{-1},$$

which is the transformation from the twice covariant compatibility matrix operator $[\partial \tilde{\partial}]$ to the twice covariant *absolute* compatibility matrix operator $[D\tilde{D}]$.

### 3.2 The *absolute* form of the Ricci matrix.

Now we introduce a new *absolute* matrix $K$, which is determined by the expression:

$$\boxed{K \equiv ([D\tilde{D}]e)e^{-1} = -e([D\tilde{D}]e^{-1})}.$$

Taking into the consideration that $e_0$ is an integrable function, we obtain also

$$K = ([D\tilde{D}]\Omega)\tilde{\Omega}.$$

The twice covariant form of the matrix $K$ is found from the matrix $K$,

$$\underline{K} = \tilde{e}\, K\, e .$$

Then we obtain

$$\underline{K} = \tilde{e}\, K\, e = \tilde{e}[D\tilde{D}]e = \tilde{e}\,\tilde{e}^{-1}\left[\partial \tilde{\partial}\right]e^{-1}\underset{\uparrow}{\underset{\downarrow}{e}} = i(\alpha)\tilde{i}(\beta)e^{-1}[\partial \tilde{\partial}]_{\alpha\beta}e = i(\alpha)\tilde{i}(\beta)\sigma^{\alpha\beta}$$

From the expressions for the elements of the matrix $\sigma^{ab}$ and the Riemann tensor $R^m_{kab}$,

$$\sigma^{ab}_{mk} = \partial_a \sigma^b_{mk} - \partial_b \sigma^a_{mk} + \sigma^a_{ms}\sigma^b_{sk} - \sigma^b_{ms}\sigma^a_{sk} = \partial_a \Gamma^m_{bk} - \partial_b \Gamma^m_{ak} + \Gamma^m_{as}\Gamma^s_{bk} - \Gamma^m_{bs}\Gamma^s_{ak}$$

$$R^m_{kab} = \partial_a \Gamma^m_{bk} - \partial_b \Gamma^m_{ak} + \Gamma^m_{as}\Gamma^s_{bk} - \Gamma^m_{bs}\Gamma^s_{ak}$$

it follows,

$$\sigma^{ab}_{mk} = R^m_{kab}$$

$$\underline{K}_{mn} = \tilde{i}(m)\underline{K}i(n) = \tilde{i}(m)i(\alpha)\tilde{i}(\beta)\sigma^{\alpha\beta}i(n) = \sigma^{m\beta}_{\beta n} = R^{\beta}_{nm\beta}$$

$$\underline{K}_{mn} = \sigma^{m\beta}_{\beta n} = \partial_m \sigma^{\beta}_{\beta n} - \partial_\beta \sigma^m_{\beta n} + \sigma^m_{\beta\lambda}\sigma^{\beta}_{\lambda n} - \sigma^{\beta}_{\beta\lambda}\sigma^m_{\lambda n}) = \partial_m \Gamma^{\beta}_{\beta n} - \partial_\beta \Gamma^{\beta}_{mn} + \Gamma^{\beta}_{m\lambda}\Gamma^{\lambda}_{\phi n} - \Gamma^{\beta}_{\beta\lambda}\Gamma^{\lambda}_{mn}$$

The last expression coincides with the definition of the Ricci matrix. Therefore we can consider the matrix $K$ as the absolute form of the Ricci matrix.

**Special notation.** For the twice covariant matrix $\underline{K}$ special notation is used,

$$R \equiv \underline{K} = i(\alpha)\tilde{i}(\beta)\sigma^{\alpha\beta} = i(\alpha)\tilde{i}(\beta)g^{-1}\gamma^{\alpha\beta}$$

The matrix $R$ is the twice covariant Ricci matrix.

### 3.3 The absolute Riemann curvature matrix.

The *absolute* two-index matrix $K^{mn} = \{K^{mn}_{ij}\}$ is determined by

$$\boxed{K^{mn} \equiv ([D\tilde{D}]_{mn}e)e^{-1} = -e([D\tilde{D}]_{mn}e^{-1})}.$$

It is easy to obtain the following equations

$$K^{mn} = (\tilde{e}^{-1} i(\alpha) \tilde{i}(\beta) e^{-1})_{mn} e \sigma^{\alpha\beta} e^{-1},$$

$$K^{\mu\nu}_{\alpha\beta} = \tilde{e}^{-1}_{\mu m} e^{-1}_{n\nu} e_{\alpha\rho} \sigma^{mn}_{\rho\gamma} e^{-1}_{\gamma\beta},$$

$$\tilde{e}_{a\alpha} \tilde{e}_{b\beta} \tilde{e}_{c\mu} \tilde{e}_{d\nu} K^{\mu\nu}_{\alpha\beta} = g_{a\rho} \sigma^{cd}_{\rho b} = g_{\underline{a}\underline{p}} R^{\overline{\rho}}_{\underline{b}\underline{c}\underline{d}} = R_{\underline{a}\underline{b}\underline{c}\underline{d}},$$

from which it follows that the matrix $K^{mn} = \{K^{mn}_{ij}\}$ is the absolute curvature matrix or the absolute matrix form of the Riemann-Christoffel tensor $R_{\underline{a}\underline{b}\underline{c}\underline{d}}$.

## 3.4 Relations between matrices and appropriate tensor expressions from the theory of relativity.

**1.** *The one-index Christoffel matrix of the first kind $\gamma^c$,*

$$\gamma^c \equiv \tilde{e} \partial_c e = \frac{1}{2}(\partial_c g + g\, i(c) \tilde{\partial} - \partial \tilde{i}(c) g), \qquad \gamma^c_{ab} = \frac{1}{2}(\partial_c g_{ab} + \partial_b g_{ac} - \partial_a g_{bc}) = \Gamma_{a,cb}.$$

**2.** *The one-index Christoffel matrix of the second kind, $\sigma^m \equiv e^{-1} \partial_m e$,*

$$\sigma^m = \frac{g^{-1}}{2}(\partial_m g + g\, i(m)\tilde{\partial} - \partial \tilde{i}(m) g), \qquad \sigma^m_{\mu\beta} = \frac{g^{-1}_{\mu\rho}}{2} g^{-1}_{\mu\rho}(\partial_\beta g_{m\rho} + \partial_m g_{\rho\beta} - \partial_\rho g_{m\beta}) = \Gamma^\mu_{m\beta}$$

**3.** Ricci matrix R,

$$R \equiv \underline{K} = i(\alpha) \tilde{i}(\beta) \sigma^{\alpha\beta} = i(\alpha) \tilde{i}(\beta) g^{-1} \gamma^{\alpha\beta}, \qquad R_{mn} = \sigma^{m\beta}_{\beta n} = R^\beta_{nm\beta},$$

$$\sigma^{m\beta}_{\beta n} = \partial_m \sigma^\beta_{\beta n} - \partial_\beta \sigma^m_{\beta n} + \sigma^m_{\beta\lambda} \sigma^\beta_{\lambda n} - \sigma^\beta_{\beta\lambda} \sigma^m_{\lambda n} = R^\beta_{nm\beta} = \partial_m \Gamma^\beta_{\beta n} - \partial_\beta \Gamma^\beta_{mn} + \Gamma^\beta_{m\lambda} \Gamma^\lambda_{\beta n} - \Gamma^\beta_{\beta\lambda} \Gamma^\lambda_{mn}$$

**4.** *The absolute Riemann curvature matrix.*

$$K^{mn} \equiv ([D\tilde{D}]_{mn} e) e^{-1} = (\tilde{e}^{-1} i(\alpha) \tilde{i}(\beta) e^{-1})_{mn} e \sigma^{\alpha\beta} e^{-1}$$

$$\tilde{e}_{a\alpha} \tilde{e}_{b\beta} \tilde{e}_{c\mu} \tilde{e}_{d\nu} K^{\mu\nu}_{\alpha\beta} = g_{a\rho} \sigma^{cd}_{\rho b} = g_{\underline{a}\underline{p}} R^{\overline{\rho}}_{\underline{b}\underline{c}\underline{d}} = R_{\underline{a}\underline{b}\underline{c}\underline{d}}$$

### 3.5 Properties of the absolute Riemann curvature and Ricci matrices.

From the integrability of the metric matrix $g = \tilde{e}\, e$ it follows

$$[D\,\tilde{D}]_{mn}(\tilde{e}\, e) = 0 = ([D\,\tilde{D}]_{mn}\tilde{e}\,)e + \tilde{e}[D\,\tilde{D}]_{mn} e = \tilde{e}\tilde{K}^{mn} e + \tilde{e} K^{mn} e ,$$

and therefore the matrix $K^{mn}$ is the antisymmetric matrix,

$$K^{mn} + \tilde{K}^{mn} = 0.$$

The integrability conditions

$$\boxed{D_j\, e^{-1}\, i(k) = D_k\, e^{-1}\, i(j), \quad j, k = 1,...,n}$$

imply the other properties of the absolute Riemann curvature matrix $K$. Using the definition of $K$ we can write

$$K^{mn} i(k) = -e\,(D_m\,(D_n\, e^{-1} i(k)) - D_n\,(D_m\, e^{-1} i(k))) = -e\,(D_m\,(D_n\, e^{-1} i(k)) - D_n\,(D_k\, e^{-1} i(m))),$$

and then we obtain

$$K^{mn} i(k) + K^{nk} i(m) + K^{km} i(n) = -e\,(D_m\,(D_n\, e^{-1} i(k)) - D_n\,(D_k\, e^{-1} i(m))) -$$
$$-e\,(D_n\,(D_k\, e^{-1} i(m)) - D_k\,(D_m\, e^{-1} i(n))) - e\,(D_k\,(D_m\, e^{-1} i(n)) - D_m\,(D_n\, e^{-1} i(k))) = 0.$$

The last expression can be written in two forms,

$$K^{mn} i(k) + K^{nk} i(m) + K^{km} i(n) = 0,$$

$$\tilde{i}(k) K^{mn} + \tilde{i}(m) K^{nk} + \tilde{i}(n) K^{km} = 0.$$

The next property of matrices $K^{mn}$ is found, using the previous property.

$$2K^{mn}_{rk} = K^{mn}_{rk} + K^{nm}_{kr} = -K^{nr}_{mk} - K^{rm}_{nk} - K^{mk}_{nr} - K^{kn}_{mr} = K^{nr}_{km} + K^{rm}_{kn} + K^{mk}_{rn} + K^{kn}_{rm}$$

$$2K^{mn}_{rk} = -K^{rk}_{nm} - K^{kn}_{rm} + K^{rm}_{kn} + K^{mk}_{rn} + K^{kn}_{rm} - K^{rk}_{nm} + K^{rm}_{kn} + K^{mk}_{rn}$$

$$2K^{mn}_{rk} = -K^{rk}_{nm} - K^{mk}_{rn} - K^{kr}_{mn} + K^{mk}_{rn} = 2K^{rk}_{mn}$$

Hence we have

$$K^{mn}_{rk} = K^{rk}_{mn}.$$

The matrix $\gamma^k$ depends only on $g$, and therefore satisfies the integrability conditions:

$$[D\tilde{D}]_{mn}\gamma^k = 0 = [D\tilde{D}]_{mn}(\tilde{e}\partial_k e) = ([D\tilde{D}]_{mn}\tilde{e})\partial_k e + \tilde{e}[D\tilde{D}]_{mn}\partial_k e,$$

From the last equation it follows

$$[D\tilde{D}]_{mn}\partial_k e = -\tilde{e}^{-1}([D\tilde{D}]_{mn}\tilde{e})\partial_k e = K^{mn}\partial_k e.$$

Using this expression, one obtains

$$[DD]_{mn}D_k e = ([D\tilde{D}]_{mn}\tilde{e}_{ks}^{-1})\partial_s e + \tilde{e}_{ks}^{-1}[D\tilde{D}]_{mn}\partial_s e =$$

$$-\tilde{K}^{mn}_{kr}\tilde{e}_{rs}^{-1}\partial_s e + \tilde{e}_{ks}^{-1}K^{mn}\partial_s e = K^{mn}_{kr}D_r e + K^{mn}D_k e,$$

and hence

$$[D\,\tilde{D}]_{mn} D_k e = K_{kr}^{mn} D_r e + K^{mn} D_k e.$$

Using the notation

$$D_{mnk} \equiv [D\,\tilde{D}]_{mn} D_k + [D\,\tilde{D}]_{nk} D_m + [D\,\tilde{D}]_{km} D_n,$$

and the equality

$$[D\,\tilde{D}]_{mn} D_k e = K_{kr}^{mn} D_r e + K^{mn} D_k e,$$

we obtain the following expressions

$$D_{mnk} e = K^{mn} D_k e + K^{nk} D_m e + K^{km} D_n e + (K_{kr}^{mn} + K_{mr}^{nk} + K_{nr}^{km}) D_r e,$$

$$[D\,\tilde{D}]_{mn} D_k e + [D\,\tilde{D}]_{nk} D_m e + [D\,\tilde{D}]_{km} D_n e = D_{mnk} e = K^{mn} D_k e + K^{nk} D_m e + K^{km} D_n e.$$

Then the following identity is written in two forms

$$D_{mnk} = D_{mnk}(1) = D_{mnk}(2),$$

where

$$D_{mnk}(1) \equiv [D\,\tilde{D}]_{km} D_n + [D\,\tilde{D}]_{mn} D_k + [D\,\tilde{D}]_{nk} D_m,$$

$$D_{mnk}(2) \equiv D_k [D\,\tilde{D}]_{mn} + D_m [D\,\tilde{D}]_{nk} + D_n [D\,\tilde{D}]_{km}.$$

For the first form

$$D_{mnk}(1)e = ([D\tilde{D}]_{mn}D_k + [D\tilde{D}]_{nk}D_m + [D\tilde{D}]_{km}D_n)e = (K^{mn}D_k + K^{nk}D_m + K^{km}D_n)e.$$

For the second form

$$D_{mnk}(2)e = D_k(K^{mn}e) + D_m(K^{nk}e) + D_n(K^{km}e) =$$
$$(D_k K^{mn} + D_m K^{nk} + D_n K^{km})e + (K^{nk}D_m + K^{mn}D_k + K^{km}D_n)e.$$

From the equality

$$D_{mnk}(1)e = D_{mnk}(2)e,$$

it follows

$$D_k K^{mn} + D_m K^{nk} + D_n K^{km} = 0,$$

or in the form for the matrix elements

$$D_k K^{mn}_{ab} + D_m K^{nk}_{ab} + D_n K^{km}_{ab} = 0.$$

The matrices $K$ and $K^{mn}$ are connected by the following contracting:

$$K_{na} = K^{\beta n}_{a\beta} = K^{n\beta}_{\beta a}.$$

The matrix element $K_{an}$ can be obtained by means of the equality $K^{mn}_{rk} = K^{rk}_{mn}$,

$$K_{an} = K^{\beta a}_{n\beta} = K^{n\beta}_{\beta a} = K_{na}.$$

Hence, the matrix $K$ is the symmetric matrix,

$$K = \tilde{K}.$$

From the equation,

$$D_k K_{ab}^{mn} + D_m K_{ab}^{nk} + D_n K_{ab}^{km} = 0,$$

with $b=k=\beta$ and $a=m=\alpha$ it follows

$$D_\beta K_{\alpha\beta}^{\alpha n} + D_\alpha K_{\alpha\beta}^{n\beta} + D_n K_{\alpha\beta}^{\beta m} = -D_\beta K_{n\beta} - D_\alpha K_{n\alpha} + D_n K_{\alpha\alpha} = 0,$$

or

$$K \underset{\uparrow\ \downarrow}{D} = D_m K\, i(m) = \frac{1}{2} D \langle K \rangle.$$

### 3.6 Properties of the absolute Riemann curvature and Ricci matrices.

We have obtained the following properties of the absolute Riemann curvature and Ricci matrices:

$$1.\ K^{mn} + \tilde{K}^{mn} = 0$$

$$2.\ K_{kr}^{mn} + K_{mr}^{nk} + K_{nr}^{km} = 0$$

.

$$3.\ K_{rk}^{mn} = K_{mn}^{rk}.$$

$$4.\ K_{a\lambda}^{\lambda n} = K_{\lambda a}^{n\lambda} = K_{na} = K_{an}.$$

$$5.\ K = \tilde{K}.$$

$$6.\ K\underset{\downarrow}{\overset{\uparrow}{D}} = D_\lambda K\, i(\lambda) = \frac{1}{2} D\langle K \rangle.$$

$$7.\ D_k K^{mn} + D_m K^{nk} + D_n K^{km} = 0.$$

## 4. THE MATHEMATICAL FIELD.

### 4.1 The absolute coordinate vector, the absolute frame matrix and the absolute coordinate space

The connection between the exact differentials $dQ$ and $dx$ is defined by the expressions:

$$dQ = Q\,\tilde{\partial}(x)\,dx = e\,dx,$$

$$e = Q\,\tilde{\partial}(x),$$

$$dx = x\,\tilde{\partial}(Q)\,dQ = e^{-1}\,dQ,$$

$$e^{-1} = x\,\tilde{\partial}(Q),$$

$$d\tilde{Q}\,dQ = d\tilde{x}\,\tilde{e}\,e\,dx,$$

$$\tilde{e}\,e = g(x),$$

where $e$ is –a non-integrable but differentiable frame matrix, $g(x)$ is the integrable metric matrix of the coordinate space $x$. We call the coordinate vector $Q$ as the *absolute* coordinate vector and the frame matrix $e$ as the *absolute* frame matrix.

The non-integrable but differentiable *absolute* frame matrix $e$ can be written in the form

$$e = \Omega e_0,$$

where $e_0$ is an integrable frame matrix and $\Omega$ is a non-integrable but differentiable orthogonal matrix,

$$\tilde{\Omega}\Omega = I_n.$$

Any $n \times n$ matrix function $F$ is integrable, if

$$(\partial_\alpha \partial_\beta - \partial_\beta \partial_\alpha) F = 0, \quad \alpha, \beta = 1,...,n.$$

Any $n \times n$ matrix function $F$ is not integrable, if

$$(\partial_\alpha \partial_\beta - \partial_\beta \partial_\alpha) F \neq 0, \quad \alpha, \beta = 1,...,n.$$

The integrability of the matrix $e_0$ means that

$$(\partial_\alpha \partial_\beta - \partial_\beta \partial_\alpha) e_0 = 0.$$

The expression for the infinitesimal invariant interval is written in the form,

$$d\tau^2 = d\tilde{Q}\ dQ = d\tilde{x}\ g(x)\ dx = d\tilde{x}\ \tilde{e}\ e\ dx = d\tilde{x}\ \tilde{e}_0\ e_0\ dx,$$

We call the space of the non-integrable but differentiable *absolute* coordinate vector $Q$ as the *absolute* coordinate space.

### 4.2 The absolute velocity vector and the absolute field matrix.
### The *absolute* (non-integrable) and coordinate (integrable) velocity vectors.

Using the notations: $U$ for the *absolute* velocity-vector and $u(x)$ for the *coordinate* velocity-vector, we can write the following definitions,

$$\boxed{u(Q) \equiv \frac{dQ}{d\tau}},$$

$$\boxed{U \equiv u(Q)},$$

$$\boxed{u(x) \equiv \frac{dx}{d\tau}},$$

There is the following connection between these two velocity vectors,

$$U = e\,u(x).$$

From the expressions for the infinitesimal invariant interval

$$\boxed{d\tau^2 = d\tilde{x}\ g\ dx = d\tilde{Q}\ dQ}.$$

it follows

$$\tilde{U}U = 1,$$

and

$$\tilde{u}(x)gu(x) = 1.$$

### 4.3 The absolute non-integrable but differentiable mathematical field matrix.

The exact differential $dU$ of the *absolute* velocity vector $U$ is written as

$$dU = Ud = (U_\uparrow \tilde{D}_\downarrow)dQ.$$

Then it is possible to write

$$\boxed{\frac{dU}{d\tau} = (U_\uparrow \tilde{D}_\downarrow)\frac{dQ}{d\tau} = (U_\uparrow \tilde{D}_\downarrow)U = PU},$$

where the *absolute* matrix $P$ is,

$$\boxed{P \equiv U_\uparrow \tilde{D}_\downarrow}.$$

We call the *absolute* matrix $P$ as the *absolute* mathematical field matrix or shortly the *absolute* field matrix.

### 4.4 Two types of absolute mathematical field matrices: the absolute symmetric and the absolute antisymmetric mathematical field matrices.

The *absolute* field matrix $P$ can be written as the sum of two matrices

$$U\underset{\uparrow}{\tilde{D}}_{\downarrow} = P = P_a + P_s,$$

where the antisymmetric *absolute* mathematical field matrix $P_a$ is

$$P_a \equiv \frac{1}{2}(U\underset{\uparrow}{\tilde{D}}_{\downarrow} - D\tilde{U}) = -\tilde{P}_a,$$

and the symmetric *absolute* mathematical field matrix $P_s$ is

$$P_s \equiv \frac{1}{2}(U\underset{\uparrow}{\tilde{D}}_{\downarrow} + D\tilde{U}) = \tilde{P}_s.$$

### 4.5 The *absolute* equation of motion.

From the equations

$$\tilde{U}U = 1,$$

$$\tilde{U}P_a U = 0,$$

$$\frac{d(\tilde{U}U)}{d\tau} = \tilde{U}(\tilde{P}_s + \tilde{P}_a)U + \tilde{U}(P_s + P_a)U = 2\tilde{U}P_s U = 0,$$

it follows that the symmetric *absolute* field matrix must satisfy the condition

$$\boxed{P_s U = 0}.$$

Therefore

$$\boxed{\dfrac{dU}{d\tau} = (U\,\tilde{D})U = (P_a + P_s)U = P_a U}.$$

We call the last equation as the *absolute* equation of motion.

### 4.6 The coordinate equation of motion.

Substituting into the *absolute* equation of motion the following expressions,

$$\dfrac{dU}{d\tau} = \dfrac{de}{d\tau} u + e \dfrac{du}{d\tau},$$

$$\dfrac{de}{d\tau} = e\,\sigma^\mu u_\mu,$$

$$P_a U = \tilde{e}^{-1} \underline{P}_a u,$$

we obtain

$$\dfrac{du}{d\tau} = -\sigma^\mu u_\mu u + g^{-1} \underline{P}_a u$$

We call the last equation as the *coordinate* equation of motion, where

$$\sigma^\mu = i(\alpha)\tilde{i}(\beta)\Gamma^\alpha_{\mu\beta}, \quad \sigma^\mu_{\beta\alpha} = \Gamma^\beta_{\mu\alpha}.$$

The scalar *coordinate* equation of motion

$$\dfrac{du_\beta(x)}{d\tau} = g^{-1}_{\beta\lambda}\underline{P}_{a\lambda\mu}\,u_\mu(x) - \sigma^\mu_{\beta\alpha}u_\mu(x)u_\alpha(x) = g^{-1}_{\beta\lambda}\underline{P}_{a\lambda\mu}\,u_\mu(x) - \Gamma^\beta_{\mu\alpha}u_\mu(x)u_\alpha(x),$$

for $P_a = 0$ becomes

$$\boxed{\frac{du_\beta(x)}{d\tau} = -\sigma^\mu_{\beta\alpha} u_\mu(x) u_\alpha(x) = -\Gamma^\beta_{\mu\alpha} u_\mu(x) u_\alpha(x)},$$

which is the differential equation of geodesic, well known from the tensor classical theory of general relativity.

### 4.7 The 4-dimensional space: the coordinate equation of motion of a charged particle in the presence of both gravitational and electromagnetic fields.

The coordinate equation of motion of a charged particle in the presence of both gravitational and electromagnetic fields is written in the form [1]:

$$\frac{du_\beta(x)}{d\tau} = \frac{e}{mc^2} F^\beta{}_\mu \, u_\mu(x) - \Gamma^\beta_{\mu\alpha} u_\mu(x) u_\alpha(x),$$

where $F^\beta{}_\mu$ is the tensor (or the coordinate matrix) of an electromagnetic field. If we compare the last equation with the scalar *coordinate* equation of motion

$$\frac{du_\beta(x)}{d\tau} = g^{-1}_{\beta\lambda} \underline{P}_{a\lambda\mu} \, u_\mu(x) - \sigma^\mu_{\beta\alpha} u_\mu(x) u_\alpha(x) = g^{-1}_{\beta\lambda} \underline{P}_{a\lambda\mu} \, u_\mu(x) - \Gamma^\beta_{\mu\alpha} u_\mu(x) u_\alpha(x),$$

we will see that the antisymmetric matrix $\vec{P}_a$,

$$\boxed{\vec{P}_a = g^{-1} \underline{P}_a},$$

is the matrix of an electromagnetic field.

If the antisymmetric field matrix $P_a$ is zero and there is only the symmetric field matrix $P_s$, the equation of motion becomes the following one:

$$\frac{du_\beta(x)}{d\tau} = -\sigma^\mu_{\beta\alpha} u_\mu(x) u_\alpha(x) = -\Gamma^\beta_{\mu\alpha} u_\mu(x) u_\alpha(x),$$

which in the four-dimensional space is the equation of motion in a gravitational field. Therefore, the symmetric field matrix $P_s$ in the four-dimensional space is the matrix of a gravitational field and the field matrix $P$ can be considered as an electromagnetic-gravitational field matrix.

## 5. THE MATHEMATICAL FIELD MATRIX IDENTITIES AND DERIVATION OF THE EQUATIONS OF THE MATHEMATICAL FIELD.

### 5.1 The absolute mathematical field matrix.

The absolute Riemann curvature and Ricci matrices give us the possibility to obtain two mathematical field matrix identities, where the absolute mathematical field matrix $P$ is defined by the expression,

$$\boxed{P \equiv U_\uparrow \tilde{D}_\downarrow}$$

Here $U$ is the absolute velocity vector and $D$ is the absolute differential operator

$$U = e\, u(x),$$

$$D = \tilde{e}^{-1} \partial.$$

**Short notation.**
In the next part of this article instead of "mathematical field" we will write shortly "field".

## 5.2 The first field matrix identitiy.

The first field matrix identity connects the absolute Ricci matrix $K$ with the *absolute field matrix P*. The coordinate vector-velocity $u(x)$ is an integrable function. Therefore

$$[D\tilde{D}]\, u(x) = 0,$$

where

$$[D\tilde{D}]_{m\mu} \equiv D_m D_\mu - D_\mu D_m,$$

$$\boxed{([D\tilde{D}]e)\, e^{-1} = K}$$

Using the last equations the expression for $\boxed{[D\tilde{D}]U}$ can be written in two forms. The first form is

$$\boxed{[D\tilde{D}]\, U = ([D\tilde{D}]\, e)\, u(x) = ([D\tilde{D}]\, e)\, e^{-1}\, U = K U},$$

and the second form is

$$\boxed{[D\tilde{D}]U = D\langle U\tilde{D}\rangle - (D\tilde{U})D = D\langle P\rangle - (\tilde{P})D}.$$

Therefore we obtain the first *absolute* field matrix identity,

$$\boxed{KU = D\langle P\rangle - (\tilde{P})D},$$

or

$$\boxed{KU = D\langle P_s\rangle + (P_a - P_s)D},$$

which has the following coordinate form

$$\underline{K}\,u(x) = R\,u(x) = \partial(x)\left\langle g^{-1}\,\underline{P_s}\right\rangle + (\tilde{\sigma}^\alpha\,(\underline{P_s}-\underline{P_a}) - \partial_\alpha(x)(\underline{P_s}-\underline{P_a}) + (\underline{P_s}-\underline{P_a})\,\sigma^\alpha)\,g^{-1}\,i(\alpha)$$

$$\underline{K}\,u(x) = R\,u(x) = \partial(x)\left\langle \overline{P_s}\,g\right\rangle - g(\sigma^\alpha\,(\overline{P_s}-\overline{P_a}) + \partial_\alpha(x)(\overline{P_s}-\overline{P_a}) + (\overline{P_s}-\overline{P_a})\,\tilde{\sigma}^\alpha)\,i(\alpha),$$

where

$$\langle P\rangle = \left\langle g^{-1}\,\underline{P_s}\right\rangle = \left\langle \overline{P_s}\,g\right\rangle,$$

$$\underline{P_s}-\underline{P_a} = i(\alpha)(\partial_\alpha\,\tilde{u}(x) + \tilde{u}(x)\,\tilde{\sigma}^\alpha)\,g.$$

$$\overline{P_s}-\overline{P_a} = g^{-1}\,i(\alpha)(\partial_\alpha\,\tilde{u}(x) + \tilde{u}(x)\,\tilde{\sigma}^\alpha)\,g.$$

We will show later, that matrix $K$ is not zero matrix only for the field of concentrated mass or charge and for nonzero density.

### 5.3 The first group of the equations of a field.

Now we will show that the absolute expression $K\,U$ can be written in the form,

$$K\,U = \rho(K)\,U,$$

where $\rho(K)$ is an eigenvalue of the *absolute* Ricci matrix. It can be done in two different cases.

In the first case all $n$ eigenvalues $\rho(K)$ are equal, satisfying the following equations

$$\rho_1(K) = \rho_2(K) = \ldots = \rho_n(K) = \rho,$$

and the metric matrix $g$ corresponds to these equations. In the second case there is such absolute velocity vector $U$, which is equal to the eigenvector $U(\rho_j(K))$,

$$U = U(\rho_j(K)).$$

In the first case the appropriate absolute and coordinate Ricci matrices satisfy the matrix equations,

$$K = \tilde{e}^{-1} \underline{K} e^{-1} = \rho I, \quad R = \underline{K} = \rho g,$$

where the twice covariant Ricci matrix $\underline{K} = R$ is connected with the absolute Ricci matrix $K$,

$$\underline{K} = \tilde{e} K e.$$

At this case we can write

$$K U = \rho U, \quad g^{-1} R u(x) = \rho u(x),$$

where

$$\rho = \rho_1(K) = \rho_2(K) = ... = \rho_n(K),$$

or

$$\rho = \rho_1(g^{-1} R) = \rho_2(g^{-1} R) = ... = \rho_n(g^{-1} R)$$

In the second case

$$K U = \mu_j U, \quad g^{-1} R u(x) = \mu_j u(x),$$

where

$$\mu_j = \rho_j(K)$$

or

$$\mu_j = \rho_j(g^{-1} R),$$

and the velocity vector $U$ is equal to the eigenvector $U(\rho_j(K))$,

$$U = U(\rho_j(K)).$$

In the both cases we have the equations

$$K U = \rho U, \quad g^{-1} R u(x) = \rho u(x).$$

In the first case we obtain the equation of the mathematical field in the form:

$$R = \rho g.$$

The last equation was considered by Arthur Stanley Eddiington as a law of gravitation. He wrote in The Mathematical Theory of Relativity: "The law $R = 0$, in empty space, is chosen by Einstein for his law of gravitation. We can suggest as an alternative to $R = 0$ the law: $R = \lambda g$, where $\lambda$ is a universal constant. There are theoretical grounds for believing that this is actually the correct form."

The first group of the *absolute* equations of the field, written in the form

$$\boxed{K U = [D \tilde{D}] U = D \langle P \rangle - (\tilde{P}) D = \rho U},$$

reminds the Maxwell equation for the electromagnetic field. For the antisymmetric absolute field it becomes the Maxwell equation of an antisymmetric field matrix $P_a$,

$$\boxed{P_a\, D = \rho_a\, U}.$$

In the 4 – dimensional space the antisymmetric field matrix becomes the electromagnetic field matrix and the last equation will be Maxwell equation for the electromagnetic field.

A. Einstein wrote [2]: "The only clue which can be drawn from experience is the vague perception that something like Maxwell's electromagnetic field has to be contained within the total field".

The coordinate form of the absolute equation of the field

$$\boxed{D\langle P\rangle - (\tilde{P})\, D = \rho U}$$

is written for the general field matrix $P$ as

$$R\, u(x) = \partial(x)\langle g^{-1}\, \underline{P_s}\rangle + (\tilde{\sigma}^\alpha\, (\underline{P_s} - \underline{P_a}) - \partial_\alpha(x)(\underline{P_s} - \underline{P_a}) +$$
$$(\underline{P_s} - \underline{P_a})\, \sigma^\alpha)\, g^{-1}\, i(\alpha) = \rho\, g\, u(x),$$

for an antisymmetrical field matrix as

$$(-\tilde{\sigma}^\alpha(g_a)\, \underline{P_a} + \partial_\alpha(x)\, \underline{P_a} - \underline{P_a}\, \sigma^\alpha(g_a))\, g_a^{-1}\, i(\alpha) = \rho_a\, g_a\, u(x)$$

and for a symmetrical field matrix as

$$\partial(x)\langle g_s^{-1}\, \underline{P_s}\rangle + (\tilde{\sigma}^\alpha(g_s)\, \underline{P_s} - \partial_\alpha(x)\, \underline{P_s} + \underline{P_s}\, \sigma^\alpha(g_s))\, g_s\, i(\alpha) = \rho_s\, g_s\, u(x).$$

### 5.4 The second field matrix identity.

The second field matrix identity connects the *absolute* Riemann curvature matrix $K^{mn}$ with the *absolute field matrix P*. The expression for $[D\tilde{D}]_{mn} U$ can be written in two forms,

$$\boxed{[D\tilde{D}]_{mn} U = K^{mn} U},$$

and

$$\boxed{[D\tilde{D}]_{mn} U = D_m\, Pi(n) - D_n\, Pi(m)}.$$

From these two expressions it follows the *second absolute field matrix identity*,

$$\boxed{K^{mn} U = D_m\, P\, i(n) - D_n\, P\, i(m)}.$$

where

$$\boxed{P = U_{\uparrow}\, \tilde{D}_{\downarrow}}.$$

The *second absolute field matrix identity* can be written also in the form:

$$K^{mn} U_k = D_m D_n U_k - D_n D_m U_k,$$

where

$$K^{mn} = (\tilde{e}^{-1} i(\alpha)\tilde{i}(\beta) e^{-1})_{mn}\, e\, \sigma^{\alpha\beta}\, e^{-1}.$$

## 5.5 The second group of the equations of the field.

Using the *absolute* identity

$$\tilde{i}(k)K^{mn} + \tilde{i}(m)K^{nk} + \tilde{i}(n)K^{km} = 0,$$

and the second *absolute* field matrix identity

$$\boxed{K^{mn} U = D_m P\, i(n) - D_n P\, i(m)}$$

we obtain the second group of the *absolute* field equations for the antisymmetric field matrix,

$$\boxed{D_m P_{a\,kn} + D_n P_{a\,mk} + D_k P_{a\,nm} = 0},$$

because

$$\tilde{i}(k)K^{mn}U + \tilde{i}(m)K^{nk}U + \tilde{i}(n)K^{km}U =$$

$$= D_m P_{kn} - D_n P_{km} + D_n P_{mk} - D_k P_{mn} + D_k P_{nm} - D_m P_{nk} =.$$

$$= 2(D_m \tilde{i}(k)P_a\, i(n) + D_n \tilde{i}(m)P_a\, i(k) + D_k \tilde{i}(n)P_a\, i(m)) = 0$$

The second group of the *absolute* field equations for the symmetric field matrix is

$$\boxed{P_s U = 0}.$$

## 5.6 The partial differential equation for the metric matrix g.

From the chapters 2 and 3 we know the following expressions,

$$R = i(\alpha)\tilde{i}(\beta)\sigma^{\alpha\beta},$$

$$\sigma^{ab} \equiv \partial_a \sigma^b - \partial_b \sigma^a + \sigma^a \sigma^b - \sigma^b \sigma^a,$$

$$\sigma^m \equiv e^{-1}\partial_m e = g^{-1}\gamma^m,$$

$$\gamma^m = \frac{1}{2}(\partial_m g + g\, i(m)\tilde{\partial} - \partial\, \tilde{i}(m)\, g).$$

The Ricci matrix is the function of the metric matrix $g$.

Using the equation of the mathematical field

$$R = \rho g,$$

for the case

$$\rho = \rho_1(K) = \rho_2(K) = ... = \rho_n(K)$$

we obtain for this case the partial differential equation for the metric matrix $g$ :

$$\rho g = i(\alpha)\tilde{i}(\beta)\sigma^{\alpha\beta},$$

where in the right part there are different partial derivatives of the metric matrix $g$. This metric matrix $g$ corresponds to the equal eigenvalues

$$\rho = \rho_1(K) = \rho_2(K) = ... = \rho_n(K)$$

of the matrix $g^{-1} R$.

## 5.7 The invariant absolute divergence of the vector $\rho U$.

The invariant absolute divergence of the vector $\rho U$ is written in the form,

$$\tilde{D}(\rho U) = \tilde{\partial} e^{-1} (e(\rho u(x))) = \tilde{i}(\lambda)(\sigma^\lambda \rho u(x) + \partial_\lambda (\rho u(x)))$$

$$\tilde{D}(\rho U) = \frac{d\rho}{d\tau} + \rho \langle P \rangle = \frac{d\rho}{d\tau} + \rho \langle \overline{P} g \rangle ,$$

For the field matrix we have

$$\rho U = K U = D \langle P \rangle - (\tilde{P}) D .$$

For the antisymmetric field matrix we obtain

$$\tilde{D}(\rho_a U) = \frac{d\rho}{d\tau} = \tilde{D}(P_a D) = 0 .$$

## 5.8 The invariant absolute and coordinate divergence of the vector $\rho U$ for the symmetric field matrix.

For the symmetric field matrix it follows from the previous equations,

$$\tilde{D}(\rho_s U) = \tilde{D}(D \langle P_s \rangle - P_s D) ,$$

,

$$\tilde{D}(\rho_s U) = (\tilde{D}\rho_s) U + \rho_s \langle P_s \rangle = \frac{d\rho_s}{d\tau} + \rho_s \langle P_s \rangle .$$

For the electromagnetic field the absolute field matrix $P_a$ is an antisymmetric matrix. Therefore the equation of continuity,

$$\tilde{D}(\rho_a U) = \frac{d\rho_a}{d\tau} = \tilde{D}(P_a D) = 0$$

is an expression of the law of charge conservation

$$\boxed{\tilde{D}(\rho_a U) = \tilde{D}(P_a D)) = \frac{d\rho_a}{d\tau} = 0}.$$

# 6. THE GENERAL SPHERICALLY SYMMETRIC SOLUTION OF THE METRIC MATRIX EQUATION FOR THE SYMMETRIC MATRIX FIELD IN THE FOUR-DIMENSIONAL METRIC SPACE.

Let us apply our matrix theory for finding the spherically symmetric solution of the metric matrix equation for the symmetric matrix field in the four-dimensional metric space. This is the case of the gravitational field. We use the "geometrized units", where all quantities which in ordinary units have dimension expressible in terms of length, time and mass, are given the dimension of a power of length. All scalars, vector and matrix elements, which we use, are in the geometrized units.

### 6.1 The "rectilinear" and "spherical" coordinate systems.

We use the "rectilinear" coordinate system with the four-dimensional vector $y$ and the corresponding "spherical" coordinate system with the four-dimensional vector $x$, where

$$y_1 = x_3 \cos x_1, \quad y_2 = x_3 \sin x_1 \sin x_2,$$
$$y_3 = x_3 \sin x_1 \cos x_2, \quad y_4 = x_4 = ct$$

Here $c$ is the speed of light, $t$ is time,

$$x_3 = \sqrt{y_1^2 + y_2^2 + y_3^2} = r, \quad x_1 = \vartheta, \quad x_2 = \varphi.$$

### 6.2 The general spherically symmetric metric matrix.

In the spherical coordinate system the metric matrix can be presented in the following form,

$$g = \{\{-h^2(x_3)\, x_3^2,\, 0,\, 0,\, 0\},\, \{0,\, -h^2(x_3)\, x_3^2 \sin^2 x_1,\, 0,\, 0\},$$
$$\{0,\, 0,\, -h^2(x_3),\, -h(x_3) f(x_3)\},\, \{0,\, 0,\, -h(x_3) f(x_3),\, \frac{1}{h^6(x_3)} - f^2(x_3)\}\}.$$

Here functions $h(x_3)$ and $f(x_3)$ are the functions which we need to find, solving the equations of the field matrices. Determinant of this metric matrix $g$ is the same as the determinant of the metric matrix of the flat space time:

$$Det(g) = -x_3^4 \sin^2 x_1.$$

We will write the metric matrix in the short form,

$$g = \{\{g_{11},\, 0,\, 0,\, 0\},\, \{0, g_{22},\, 0,\, 0\}, \{0,\, 0,\, g_{33},\, g_{34}\},\, \{0,\, 0, g_{34},\, g_{44}\}\},$$

where

$$g_{11} = -h^2(x_3) x_3^2, \quad g_{22} = -h^2(x_3) x_3^2 \sin^2 x_1, \quad g_{33} = -h^2(x_3),$$

$$g_{34} = -h(x_3) f(x_3), \quad g_{44} = \frac{1}{h^6(x_3)} - f^2(x_3)$$

### 6.3 The 4-velocities.

The solution of the equations of motion (geodesic equations for the symmetrical matrix field),

$$\boxed{\frac{du_\beta(x)}{d\tau} = -\sigma^\mu_{\beta\alpha} u_\mu(x) u_\alpha(x) = -\Gamma^\beta_{\mu\alpha} u_\mu(x) u_\alpha(x)}$$

gives the expressions for 4-velocities,

$$u_1(x) = -\frac{\sqrt{c_1^2 - c_2^2 \cot^2 x_1}}{g_{11}}$$

$$u_2(x) = -\frac{c_2}{g_{22}}$$

$$u_3(x) = c_3 \sqrt{g_{44} \, g_{33} \left(\frac{c_1^2 + c_2^2}{x_3^2} - g_{33}\right) + c_4^2 \, g_{33}^{\,2}}$$

$$u_4(x) = -\frac{c_4 + g_{34} u_3(x)}{g_{44}}.$$

Here $c_1, c_2, c_3$ and $c_4$ are constants, where

$$c_3 = \pm 1.$$

## 6.4 The eigenvalues of the matrix $g^{-1}R$.

Using the short notations,

$$h \equiv h(x_3), \quad f \equiv f(x_3),$$

we will find the four eigenvalues $\mu_1, \mu_2, \mu_3, \mu_4$ of the matrix $g^{-1}R$,

$$\mu_1 = \frac{1}{x_3 h^4}\left( x_3^3 h^8 (f')^2 - 9 x_3 (h')^2 + x_3 f^2 h^7 (3h + 4x_3 h') + x_3^2 f h^7 (2f'(3h + 2x_3 h') + x_3 h f'') + 3h(2h' + x_3 h'') \right)$$

$$\mu_2 = \frac{1}{x_3 h^4}\left( x_3^3 h^8 (f')^2 - 13 x_3 (h')^2 + x_3^2 f h^7 (2f'(3h + 2x_3 h') + x_3 h f'') + h(-2h' + x_3 h'') + x_3 f^2 h^6 (3h^2 + 4x_3^2 (h')^2 + 2x_3 h(6h' + x_3 h'')) \right)$$

$$\mu_3 = \mu_4 = \frac{1}{x_3^2 h^4}\left( 3(h')^2 + 2x_3 f h^7 f'(h + x_3 h') - h h'' + f^2 h^6 (3h^2 + 3x_3^2 (h')^2 + x_3 h(8h' + x_3 h'')) \right)$$

The 4-velocity vector $u(x)$ will be the eigenvector for the equal eigenvalues

$$\mu_1 = \mu_2 = \mu_3 = \mu_4 = \rho.$$

For the difference $\mu_1 - \mu_2$ we obtain the following expression,

$$\mu_1 - \mu_2 = -\frac{2(-1 + x_3^2 f^2 h^6)(2 x_3 (h')^2 + h(4h' + x_3 h''))}{x_3 h^4}.$$

It is possible to check, that the function

$$h = (1 - (\frac{c_5}{x_3})^3)^{1/3} = \frac{(x_3^3 - c_5^3)^{1/3}}{x_3},$$

satisfies the differential equation

$$\mu_1 - \mu_2 = 0.$$

Using the above expression for the function $h$ we find the expression $\mu_2 - \mu_3$:

$$\mu_2 - \mu_3 = \frac{1}{x_3^{12}(1 - (\frac{c_5}{x_3})^3)^{8/3}} (-10 x_3^4 c_5^3 (x_3^3 + c_5^3) + (x_3^3 - c_5^3)^2 (-c_5^6 f^2 + x_3^2 (x_3^3 - c_5^3)^2 (f')^2 +$$

$$+ x_3 f (2 (2 x_3^6 - 3 x_3^3 c_5^3 + c_5^6) f' + x_3 (x_3^3 - c_5^3)^2 f'')))$$

,

The solution of the differential equation

$$\mu_2 - \mu_3 = 0,$$

gives the following function $f$:

$$f = \frac{\sqrt{(c_5^3 (2 x_3^3 - c_5^3) + \frac{2 c_6}{3}(x_3^3 - c_5^3)^{5/3} + c_7 (x_3^3 - c_5^3)^{8/3})}}{x_3 (x_3^3 - c_5^3)}.$$

In the last expression $c_5, c_6, c_7$ are constants. For the obtained functions $h$ and f we find the equal eigenvectors,

$$\mu_1 = \mu_2 = \mu_3 = \mu_4 = 3 c_7 = \rho.$$

Therefore

$$c_7 = \frac{\rho}{3},$$

where $\rho$, cosmological constant, has the dimension of density and we have the following expression for the Ricci matrix:

$$R = \rho g.$$

Now we can write the following expressions:

$$g = \{\{g_{11}, 0, 0, 0\}, \{0, g_{22}, 0, 0\}, \{0, 0, g_{33}, g_{34}\}, \{0, 0, g_{34}, g_{44}\}\},$$

where

$$g_{11} = -h^2(x_3) x_3^2, \quad g_{22} = -h^2(x_3) x_3^2 \sin^2 x_1, \quad g_{33} = -h^2(x_3),$$

$$g_{34} = -h(x_3) f(x_3), \quad g_{44} = \frac{1}{h^6(x_3)} - f^2(x_3)$$

$$h = (1 - (\frac{c_5}{x_3})^3)^{1/3} = \frac{(x_3^3 - c_5^3)^{1/3}}{x_3}$$

$$f = \frac{c_8(c_5^3(2x_3^3 - c_5^3) + \frac{2c_6}{3}(x_3^3 - c_5^3)^{5/3} + c_7(x_3^3 - c_5^3)^{8/3})^{1/2}}{x_3(x_3^3 - c_5^3)}$$

$$g_{44} = 1 - \frac{2c_6}{3 x_3 h} - c_7 x_3^2$$

The constants of integration can be determined from the motion. The dimension of $c_6$ is a length, which is proportional to the gravitational radius of the rest mass $M$ of the gravitating body and is written as

$$\frac{c_6}{3} = r_M,$$

where $r_M$ is the gravitational radius of the rest mass $M$ of the gravitating body,

$$r_M \equiv \frac{G_g M}{c^2},$$

and $G_g$ is the gravitational constant. The constant $c_5$ is proportional to the gravitational radius $r_m$,

$$r_m \equiv \frac{G_g m}{c^2},$$

of the rest mass *m* of the body, moving in the gravitational field of the rest mass *M*.

### 6.5 The coordinate cotravariant–covariant Riemann curvature matrix $\sigma^{ab}$.

Let us introduce the following notations:

$$w_0 = x_3^3 - c_5^3,$$

$$w_1 = \frac{-c_6 + 3 w_0 c_7}{3 w_0^{1/3}},$$

$$w_2 = \frac{-2 c_6 + 3 w_0 c_7}{3 w_0^{1/3}},$$

$$w_3 = \frac{c_8 (2 x_x^3 c_5^3 - c_5^6 + \frac{2}{3}(w_0^{5/3} c_6 + w_0^{8/3} c_7)^{1/2}}{x_3 w_0^{4/3}},$$

$$w_4 = \frac{2 c_6 - 3(w_0^{1/3} - w_0 c_7)}{3 w_0},$$

$$w_5 = \frac{c_8}{x_3 w_0^{4/3}}((-\frac{2}{3} c_5^3 w_0^{2/3} c_6 + x_3^6 w_0^{2/3} c_7 + c_5^6 (-1 + w_0^{2/3} c_7 + x_3^3 (\frac{2}{3} w_0^{2/3} c_6 + 2 c_5^3 (1 - w_0^{2/3} c_7))^{1/2}$$

With these notations the metric matrix *g* is written by the following way:

$$g = \{\{g_{11}, 0, 0, 0\},\ \{0, g_{22}, 0, 0\}, \{0, 0, g_{33}, g_{34}\},\ \{0, 0, g_{34}, g_{44}\}\},$$

where

$$g_{11} = -w_0^{2/3}, \quad g_{22} = -w_0^{2/3} \sin^2 x_1, \quad g_{33} = -\frac{w_0^{2/3}}{x_3^2},$$

$$g_{34} = -w_3 \, w_0^{2/3}, \quad g_{44} = 1 + w_2.$$

Using this metric matrix, we can obtain the two-index coordinate cotravariant–covariant *Riemann curvature matrices* $\sigma^{ab}$,

$$\sigma^{12} = w_2 \{\{0, -\sin^2 x_1, 0, 0\}, \{1, 0, 0, 0\}, \{0, 0, 0, 0\}, \{0, 0, 0, 0\}\},$$

$$\sigma^{13} = w_1 \{\{0, 0, \frac{1}{x_3^2}, w_3\}, \{0, 0, 0, 0\}, \{-1, 0, 0, 0\}, \{0, 0, 0, 0\}\},$$

$$\sigma^{14} = w_1 \{\{0, 0, w_3, w_4\}, \{0, 0, 0, 0\}, \{0, 0, 0, 0\}, \{-1, 0, 0, 0\}\},$$

$$\sigma^{23} = w_1 \{\{0, 0, 0, 0\}, \{0, 0, \frac{1}{x_3^2}, w_3\}, \{0, -\sin^2 x_1, 0, 0\}, \{0, 0, 0, 0\}\},$$

$$\sigma^{24} = w_1 \{\{0, 0, 0, 0\}, \{0, 0, w_3, w_4\}, \{0, 0, 0, 0\}, \{0, -\sin^2 x_1, 0, 0\}\},$$

$$\sigma^{34} = w_2 \{\{0, 0, 0, 0\}, \{0, 0, 0, 0\}, \{0, 0, -w_5, -w_4\}, \{0, 0, \frac{1}{x_3^2}, w_5\}\}.$$

The $n$ – dimensional space becomes flat if all elements of the two-index coordinate cotravariant–covariant *Riemann curvature matrix* $\sigma^{ab}$ become zero. From the last six expressions it follows that the *Riemann curvature matrix* $\sigma^{ab}$ becomes zero only if

$$w_1 = \frac{-c_6 + 3 w_0 c_7}{3 w_0^{1/3}} = 0,$$

and

$$w_2 = \frac{-2c_6 + 3w_0 c_7}{3w_0^{1/3}} = 0,$$

or

$$c_6 = 0$$

and

$$c_7 = 0.$$

The curvature of the space is defined by the constants $c_6$ and $c_7$. The constant $c_6$ is the concentrated mass in the units of length. This constant is proportional to the gravitational radius of the rest mass $M$ of the gravitating body. The constant $c_7$ is proportional to the cosmological density.

We have found a rigorous nonsingular spherically symmetric solution of the gravitational equations which may be applied to the investigation of the motion of planets around the Sun. At this case the mass $M$ is the rest mass of the Sun. The second constant $c_7$ can be determined from the measurements of the precession of perihelia. For the motion of Mercury we obtain $c_7 = \frac{\rho}{3} \approx 1.3*10^{-33}\ km^{-2}$ if the major axis of Mercury's orbit precesses at a rate of 43.1 arcsecs every 100 years.

### 6.6 The equations for the radius $x_3$ as a function of $x_2$.

For the plane orbital motion, where

$$y_1 = const = y_{10} = \frac{\pi}{2}, \quad c_1 = 0,$$

the radius $x_3$ as a function of $x_2$ for the general spherically symmetric metric matrix is determined by the differential equation

$$\frac{d^2k}{dx_2^2} = \frac{1}{3kc_2^2}(-3k^2 c_2^2 - 3d(1-k^3 c_5^3)^{2/3}c_7 + 3k^5 c_2^2(dc_6 + c_5^3(5-4dc_7)) +$$
$$4k^8 c_2^2 c_5^3(-2dc_6 + 3c_5^3(-1+dc_7)) + 4k^6 c_5^3(1-k^3 c_5^3)^{2/3}(-2dc_6 + 3c_5^3(-1+c_4^2+dc_7)) +$$
$$k^3(1-k^3 c_5^3)^{2/3}((dc_6 - 3c_5^3(-4+4c_4^2+3dc_7))),$$

$$k \equiv \frac{1}{x_3},$$

$$d \equiv (\frac{1}{k^3} - c_5^3)^{2/3}.$$

For the particular case

$$c_5 = 0, \quad h = 1, \quad c_7 = 0, \quad d = \frac{1}{k^2},$$

$$\frac{c_6}{3} = r_M,$$

we obtain

$$f = c_8 \frac{\sqrt{\frac{2r_M}{x_3}}}{x_3},$$

$$\frac{d^2k}{dx_2^2} = -k + 3k^2 r_M + \frac{r_M}{c_2^2},$$

$$g = \{\{g_{11}, 0, 0, 0\}, \{0, g_{22}, 0, 0\}, \{0, 0, g_{33}, g_{34}\}, \{0, 0, g_{34}, g_{44}\}\}$$

$$g_{11} = -x_3^2, \quad g_{22} = -x_3^2 \sin^2 x_1, \quad g_{33} = -1,$$

$$g_{34} = -c_8 \sqrt{\frac{2r_M}{x_3}}, \quad g_{44} = 1 - \frac{2r_M}{x_3}$$

## 7. THE GENERAL SPHERICALLY SYMMETRIC WEAK SOLUTION FOR $c_5 = 0$.

### 7.1 The metric matrix and 4-velocities.

Let us consider the particular case of the general spherically symmetric solution for $c_5 = 0$, $h = 1$. For this case of the weak gravitational force we obtain the following expressions:

$$g = \{\{-x_3^2, 0, 0, 0\}, \{0, -x_3^2 (\sin x_1)^2, 0, 0\},$$

$$\{0, 0, -1, -\sqrt{\frac{2r_M}{x_3} + c_7 x_3^2}\}, \{0, 0, -\sqrt{\frac{2r_M}{x_3} + c_7 x_3^2}, 1 - \frac{2r_M}{x_3} - c_7 x_3^2\}$$

$$R = 3c_7 g = \rho g,$$

$$u_1(x) = \frac{\sqrt{c_1^2 - c_2^2 \cot^2 x_1}}{x_3^2},$$

$$u_2(x) = \frac{c_2}{(x_3 \sin x_1)^2},$$

$$u_3(x) = \sqrt{c_4^2 - (1 - \frac{2\,r_M}{x_3} - c_7\, x_3^2)(\frac{c_1^2 + c_2^2}{x_3^2} + 1)},$$

$$u_4(x) = -\frac{c_4 - u_3(x)\sqrt{\frac{2\,r_M}{x_3} + c_7\, x_3^2}}{1 - \frac{2\,r_M}{x_3} - c_7\, x_3^2}.$$

For the particular case of the zero density $\rho$,

$$c_7 = 0,$$

we obtain

$$f = \frac{\sqrt{\frac{2\,r_M}{x_3}}}{x_3},$$

$$g = \{\{g_{11}, 0, 0, 0\},\ \{0, g_{22}, 0, 0\}, \{0, 0, g_{33}, g_{34}\}, \{0, 0, g_{34}, g_{44}\}\}$$

$$g_{11} = -x_3^2, \quad g_{22} = -x_3^2 \sin^2 x_1, \quad g_{33} = -1,$$

$$g_{34} = -\sqrt{\frac{2 r_M}{x_3}}, \quad g_{44} = 1 - \frac{2 r_M}{x_3}$$

## 7.2 The general form of spherically symmetric solution for $c_5 = 0$ (weak solution) with one arbitrary function.

It is possible to show that the general form of spherically symmetric solution for $c_5 = 0$ (weak solution) is

$$\begin{aligned} g = \{&\{-x_3^2,\, 0,\, 0,\, 0\},\, \{0,\, -x_3^2 \sin^2 x_1,\, 0,\, 0\}, \\ &\{0,\, 0,\, g_{33}(x_3),\, c_8 \sqrt{1 + g_{33}(x_3) g_{44}(x_3)}\}, \\ &\{0,\, 0,\, c_8 \sqrt{1 + g_{33}(x_3) g_{44}(x_3)},\, g_{44}(x_3)\}\} \end{aligned}$$

where $g_{33}(x_3)$ is an arbitrary function, and

$$g_{44} = 1 - \frac{2 r_M}{x_3} - \frac{\rho}{3} x_3^2.$$

This form implies that the radius vector $x_3$ is defined in such a way that the circumference of a circle with center at the origin of coordinates is equal to $2\pi x_3$ (the element of arc of a circle in the plane $x_1 = \pi/2$ is equal to $ds = x_3\, dx_2$) [1]. The coefficient $c_8 = 1$ for $x_4 > 0$ and $c_8 = -1$ for $x_4 < 0$. At this case the term $g_{34} dx_3\, dx_4 + g_{43}\, dx_4\, dx_3$) will be invariant under the time reversal.

For this metric matrix we have

$$R = \rho g.$$

### 7.3 The Schwarzschild solution.

The Schwarzschild diagonal metric matrix $g_s$ is the particular case of the general weak metric matrix, where the function $g_{33}(x_3)$ is taken in the form:

$$g_{33}(x_3) = -\frac{1}{g_{44}(x_3)} = -\frac{1}{1 - \frac{2r_M}{x_3} - \frac{\rho}{3}x_3^2}.$$

For this function $g_{33}(x_3)$ and for $\rho = 0$ we obtain the Schwarzschild metric matrix $g_S$,

$$g = g_S \equiv \{\{-x_3^2, 0, 0, 0\}, \{0, -(x_3 \sin x_1)^2, 0, 0\},$$
$$\{0, 0, -\frac{1}{1 - \frac{2r_M}{x_3}}, 0\}, \{0, 0, 0, 1 - \frac{2r_M}{x_3}\}\},$$

which has a singularity at the Schwarzschild radius $x_3 = 2r_M$. S. Hawking comments this singularity [3]: "However, this is just caused by a bad choice of coordinates. One can choose other coordinates in which the metric is regular there."

### 7.4 The 4-velocities for the general weak metric matrix with $\rho = 0$.

For the general weak metric with $\rho = 0$,

$$g = \{\{-x_3^2, 0, 0, 0\}, \{0, -x_3^2 \sin^2 x_1, 0, 0\}, \{0, 0, -1, -\sqrt{\frac{2r_M}{x_3}}\}, \{0, 0, -\sqrt{\frac{2r_M}{x_3}}, 1 - \frac{2r_M}{x_3}\}\}$$

we find

$$u_1(x) = -\frac{\sqrt{c_1^2 - c_2^2 \cot^2 x_1}}{x_3^2},$$

$$u_2(x) = \frac{c_2}{y_3^2 \sin^2 x_1},$$

$$u_3(x) = c_3\sqrt{c_4^2 - (1 - \frac{2r_M}{x_3})(\frac{c_1^2 + c_2^2}{x_3^2} + 1)}, \quad c_3 = \pm 1,$$

$$u_4(x) = -\frac{c_4 - u_3(x)\sqrt{\frac{2r_M}{x_3}}}{1 - \frac{2r_m}{x_3}}.$$

Then dimensionless velocities become

$$\beta_1(x_3) = \frac{u_1(x) x_3}{u_4(x)}, \quad \beta_2(x_3) = \frac{u_2(x) x_3}{u_4(x)}, \quad \beta_3(x_3) = \frac{u_3(x)}{u_4(x)}.$$

### 7.5 The plane orbital motion.

For the plane orbital motion with $\rho = 0$, where

$$x_1 = const = \frac{\pi}{2}, \quad c_1 = 0, \quad u_1(x) = 0,$$

the constants are expressed through the perihelion distance $p$ and the aphelion distance $a$, taking into consideration that $u_3(x) = 0$ at $x_3 = p$ and at $x_3 = a$,

$$c_2 = -\frac{a\,p\sqrt{2r_M}}{f_2}, \quad c_4 = \frac{f_3}{f_2},$$

$$f_2 = \sqrt{a\,p\,(a+p) - 2\,r_m\,(a^2 + a\,p + p^2)},$$

$$f_3 = \sqrt{(a+p)(a - 2\,r_M)(p - 2\,r_M)} = \sqrt{a\,p\,(a+p)\,h(a)\,h(p)}.$$

Then we have

$$u_2(x) = \frac{\sqrt{2\,r_M}\,a\,p}{f_2\,x_3^2},$$

$$u_3(x) = \frac{\sqrt{2\,r_M}\,f_1}{f_2\,x_3^{3/2}},$$

$$u_4(x) = \frac{f_3\,x_3}{f_2\,(x_3 - 2\,r_M)} + \frac{2\,r_M\,f_1}{(x_3 - 2\,r_M)\,f_2\,x_3}.$$

with

$$f_1 = \sqrt{(a - x_3)(x_3 - p)(a\,(p\,x_3 - 2\,r_M\,(x_3 + p)) - 2\,r_M\,p\,x_3)}.$$

For the Schwarzschild model

$$u_4(x) = \frac{f_3\,x_3}{f_2\,(x_3 - 2\,r_M)}$$

This is the only difference between our model and the Schwarzschild model. The extreme dimensionless transverse velocities are the same for the both metrics and they are written in the form:

$$\beta_{2\max} = -\beta_2(a) = -\frac{a}{p}\sqrt{\frac{2r_M(p-2r_M)}{(a+p)(a-2r_M)}},$$

$$\beta_{2\min} = -\beta_2(p) = -\frac{p}{a}\sqrt{\frac{2r_M(a-2r_M)}{(a+p)(p-2r_M)}}.$$

The radius $x_3$ as a function of $x_2$ for the both models is determined by the same differential equation

$$\frac{d^2k}{dx_2^2} = -k + 3k^2 r_M + \frac{r_M}{c_2^2}, \quad k = \frac{1}{x_3}.$$

Let $x_{21}$ be the position of perihelion and $x_{22}$ is the position of perihelion after one period of revolution. Using a method of successive approximation, we find the difference $\Delta\varphi$, which gives the displacement of perihelion after one period of revolution,

$$\Delta\varphi = x_{22} - x_{21} - 2\pi = \frac{3r_M f_4}{f_5},$$

where

$$f_4 = a^2 p^2 (a^2 - p^2) + 4(a^2 + ap + p^2) r_M (ap^2 - r_M (a^2 + ap + p^2)),$$

$$f_5 = a^2 p^2 (a-p)(ap - 2r_M(2a+p)).$$

The Table 1 shows the precession of planets as the results of calculations using the expression

$$\Delta\varphi = x_{22} - x_{21} - 2\pi = \frac{3 r_M f_4}{f_5}.$$

Table 1.

Relativistic precession (arc sec) of planets.

|  | Precession per rev formula | Precession per century observed | Precession per century formula |
|---|---|---|---|
| Mercury | 0.103517 | 43.1±0.5 | 42.9531 |
| Venus | 0.0530579 | 8.4±4.8 | 8.6273 |
| Earth | 0.0383884 | 5.0±1.2 | 3.83884 |
| Mars | 0.0254106 |  | 1.35091 |
| Jupiter | 0.00739156 |  | 0.0623129 |
| Saturn | 0.0040177 |  | 0.0136392 |
| Uranus | 0.00200285 |  | 0.00238403 |
| Neptune | 0.00127736 |  | 0.000775147 |
| Pluto | 0.00104024 |  | 0.000419995 |

In the Table 2 we show the calculated extreme velocities of planets, using the formulae

$$\beta_{2\max} = -\beta_2(a) = -\frac{a}{p}\sqrt{\frac{2 r_M (p - 2 r_M)}{(a+p)(a - 2 r_M)}},$$

$$\beta_{2\min} = -\beta_2(p) = -\frac{p}{a}\sqrt{\frac{2r_M(a-2r_M)}{(a+p)(p-2r_M)}}.$$

and the NASA data of the perihelion and aphelion distances.

Table 2.

Calculated extreme orbital velocities (km/s) of planets for the NASA data of perihelion and aphelion distances.

|  | $v_{min}$ Nasa data | $v_{min}$ Formula | $v_{max}$ Nasa data | $v_{max}$ Formula |
|---|---|---|---|---|
| Mercury | 38.86 | 38.8568 | 58.98 | 58.9779 |
| Venus | 34.79 | 34.7850 | 35.26 | 35.2575 |
| Earth | 29.29 | 29.2903 | 30.29 | 30.2879 |
| Mars | 21.97 | 21.9708 | 26.50 | 26.5017 |
| Jupiter | 12.44 | 12.4327 | 13.72 | 13.7103 |
| Saturn | 9.09 | 9.0927 | 10.18 | 10.1815 |
| Uranus | 6.49 | 6.49358 | 7.11 | 7.11496 |
| Neptune | 5.37 | 5.37276 | 5.50 | 5.49512 |
| Pluto | 3.71 | 3.70514 | 6.10 | 6.10229 |

## 7.6 The case of purely radial motion.

For the case of purely radial motion

$$\beta_{10} = 0, \quad \beta_{20} = 0, \quad y_1 = \frac{\pi}{2}.$$

In the Schwarzschild geometry the dimensionless radial velocity becomes

$$\beta_3 = \pm \frac{y_3 - 2r_M}{c_4 y_3} \sqrt{c_4^2 - 1 + \frac{2r_M}{y_3}}.$$

The radial velocity reaches an extreme value $\beta_{3m}$,

$$|\beta_{3m}| = \frac{2c_4^2}{3\sqrt{3}},$$

at the radius $y_{3m}$, where

$$y_{3m} = \frac{2r_M}{1 - \frac{4c_4^4}{9}},$$

$$c_4 = (1 - \frac{2r_M}{y_{30}}) \sqrt{\frac{1 - \frac{2r_M}{y_{30}}}{(1 - \frac{2r_M}{y_{30}})^2 - \beta_{30}^2}}.$$

If the particle begins accelerating at large radius ($y_3 \to \infty$) with zero initial velocity $\beta_{30}$ ($c_4 \to 1$), it will attain a maximum value $\beta_{3\max}$,

$$|\beta_{3\max}| = \frac{2}{3\sqrt{3}} = 0.39.$$

For

$$y_3 \to 2r_M, \quad \beta_3 \to 0.$$

We have obtained the known result that for the Schwarzschild geometry, in the region $2r_M < y_3 < \infty$, the velocity of the particle undergoes variations unlike the monotonic increase predicted by Newtonian gravitational theory.

In our geometry the dimensionless radial velocity becomes

$$\beta_{3r}(y_3) = \frac{(1 - \frac{2r_M}{y_3})\sqrt{\alpha_{4r}^2 - (1 - \frac{2r_M}{y_3})}}{\sqrt{\frac{2r_M}{y_3}(\alpha_{4r}^2 - (1 - \frac{2r_M}{y_3}))} \pm \alpha_{4r}},$$

with

$$\alpha_{4r} = \frac{1 - \frac{2r_M}{y_{30}} - \beta_{30}\sqrt{\frac{2r_M}{y_{30}}}}{\sqrt{1 - \frac{2r_M}{y_{30}} - 2\beta_{30}\sqrt{\frac{2r_M}{y_{30}}} - \beta_{30}^2}}.$$

For the particle which begins accelerating with zero initial velocity $\beta_{30}$ we find for the negative velocity

$$\alpha_{4r0} = \sqrt{1 - \frac{2r_M}{y_{30}}},$$

$$\beta_{3r0}(y_3) = \frac{(1-\frac{2r_M}{y_3})\sqrt{1-\frac{1-\frac{2r_M}{y_3}}{1-\frac{2r_M}{y_{30}}}}}{\sqrt{\frac{2r_M}{y_3}\left(1-\frac{1-\frac{2r_M}{y_3}}{1-\frac{2r_M}{y_{30}}}\right)-1}}.$$

For the particle which begins accelerating at large radius ($y_{30} \to \infty$) with zero initial velocity $\beta_{30}$ we find for the negative velocity

$$\alpha_{4r0} \to 1, \quad \beta_{3r0}(y_3) \to -\sqrt{\frac{2r_M}{y_3}}.$$

For

$$y_3 \to 2r_M, \quad \beta_{3r0}(y_3) \to -1.$$

We see that in our geometry the velocity of the particle undergoes the monotonic increase predicted by the Newtonian gravitational theory.

.

## 8. THE GENERAL SPHERICALLY SYMMETRIC SOLUTION OF THE METRIC MATRIX EQUATION FOR THE SYMMETRIC MATRIX FIELD IN THE FOUR-DIMENSIONAL METRIC SPACE AND RECTLINEAR COORDINATE SYSTEM.

### 8.1 The "rectilinear" coordinate system.

We use the "rectilinear" coordinate system with the four-dimensional vector *x*, where

$$x_4 = ct,$$

$$r = \sqrt{x_1^2 + x_2^2 + x_3^2}$$

**8.2 The spherically symmetric metric matrix in the rectilinear coordinate system.**

In the rectlinear coordinate system the metric matrix can be presented in the following form,

$$g = \{\{-w(r), 0, 0, f(r)x_1\}, \{-w(r), 0, 0, f(r)x_2\},$$
$$\{-w(r), 0, 0, f(r)x_3\}, \{f(r)x_1, f(r)x_2, f(r)x_3, \frac{1}{w^3(r)} - \frac{r^2 f^2(r)}{w(r)}\}\}.$$

Here functions $w(r)$ and $f(r)$ are the functions which we need to find, solving the equations of the field matrices. Determinant of this metric matrix $g$ is the same as the determinant of the metric matrix of the flat space time:

$$Det(g) = -1.$$

We will write the metric matrix in the short form,

$$g = \{\{g_{11}, 0, 0, g_{14}\}, \{0, g_{22}, 0, g_{24}\}, \{0, 0, g_{33}, g_{34}\}, \{g_{14}, g_{24}, g_{34}, g_{44}\}\},$$

where

$$g_{11} = g_{22} = g_{33} = -w(r), \quad g_{14} = f(r)x_1, \quad g_{24} = f(r)x_2,$$

$$g_{34} = f(r)x_3, \quad g_{44} = \frac{1}{w^3(r)} - \frac{r^2 f^2(r)}{w(r)}$$

**8.3 The eigenvalues of the square matrix $g^{-1}R$.**

Using the short notations,

$$w \equiv w(r), \quad f \equiv f(r),$$

we will find the four eigenvalues $\mu_1, \mu_2, \mu_3, \mu_4$ of the square matrix $g^{-1}R$,

$$\mu_1 = \mu_2 = \frac{1}{2w^3}(2(w')^2 + 2rfw^3 f'(2w+rw') - ww'' + f^2 w^3(6w+r(6w'+rw'')))$$

$$\mu_3 = \frac{1}{2rw^3}(2r^3 w^4 (f')^2 - 7r(w')^2 + 2r^2 fw^4(6f'+rf'') + w(-2 w'+rw''+$$
$$rf^2 w^2 (6w^2 + r^2 (w')^2 + rw(6w'+rw'')))$$

$$\mu_4 = \frac{1}{2rw^3}(2r^3 w^4 (f')^2 - 6r(w')^2 + 2r^2 fw^4(6f'+rf'') + 3w(2 w'+rw''-$$
$$rf^2 w^3 (-6w+r(2 w'+rw''))))$$

The 4-velocity vector $u(x)$ will be the eigenvector for the equal eigenvalues

$$\mu_1 = \mu_2 = \mu_3 = \mu_4 = \rho.$$

For the difference $\mu_3 - \mu_4$ we obtain the following expression,

$$\mu_3 - \mu_4 = \frac{(-1 + r^2 f^2 w^2)(r(w')^2 + 2w(4w'+rw''))}{2rw^3}.$$

It is possible to check, that the function

$$w = \frac{(r^3 - c_5^3)^{2/3}}{r^2},$$

satisfies the differential equation

$$\mu_3 - \mu_4 = 0.$$

Using the above expression for the function $w$ we find the expression $\mu_2 - \mu_3$:

$$\mu_2 - \mu_3 = \frac{1}{r^2(r^3 - c_5^3)^{8/3}}(10\, r^2\, c_5^3(r^3 + c_5^3) - (r^3 - c_5^3)^{4/3}(2c_5^6\, f^2 + r^2(r^3 - c_5^3)^2(f')^2 +$$
$$+ r(r^3 - c_5^3)f((4r^3 - 6c_5^3)f' + r(r^3 - c_5^3)f'')))$$

,

The solution of the differential equation

$$\mu_2 - \mu_3 = 0,$$

gives the following function $f$,

$$f = \frac{1}{r^2}\sqrt{(r^3 - c_5^3)^{1/3}(c_5^3\frac{(2r^3 - c_5^3)}{(r^3 - c_5^3)^{5/3}} + (2c_6 + r^3\, c_7))}$$

In the last expression $c_5, c_6, c_7$ are constants. For the obtained functions $h$ and $f$ we find the equal eigenvectors,

$$\mu_1 = \mu_2 = \mu_3 = \mu_4 = 3c_7 = \rho.$$

Therefore

$$c_7 = \frac{\rho}{3},$$

where $\rho$, cosmological constant, has the dimension of density and we have the following expression for the Ricci matrix:

$$R = \rho g.$$

Now we can write the following expressions:

$$w = \frac{(r^3 - c_5^3)^{2/3}}{r^2},$$

$$f = \frac{1}{r^2}\sqrt{(r^3 - c_5^3)^{1/3}(c_5^3 \frac{(2r^3 - c_5^3)}{(r^3 - c_5^3)^{5/3}} + (2c_6 + r^3 c_7))}$$

$$g_{44} = 1 - \frac{2c_6 + r^3 c_7}{(r^3 - c_5^3)^{1/3}}$$

The constants of integration can be determined from the motion. The dimension of $c_6$ is a length, which is proportional to the gravitational radius of the rest mass $M$ of the gravitating body and is written as

$$\frac{c_6}{3} = r_M,$$

where $r_M$ is the gravitational radius of the rest mass $M$ of the gravitating body,

$$r_M \equiv \frac{G_g M}{c^2},$$

and $G_g$ is the gravitational constant. The constant $c_5$ is proportional to the gravitational radius $r_m$,

$$r_m \equiv \frac{G_g m}{c^2},$$

of the rest mass $m$ of the body, moving in the gravitational field of the rest mass $M$.

## 9. THE RADIUS OF MASS AND THE RADIUS OF CHARGE OF CHARGED MASS.

For the gravitational field of mass the radius of mass (the mass in the units of length) is

$$r_M \equiv \frac{G_g M}{c^2},$$

where $G_g$ is the gravitational constant and $M$ is the central mass.
The mass of the sun, $1.99 \cdot 10^{33}$ grams, becomes in the units of length 1.47 kilometers, and other masses are converted in a like proportion. One gram in the units of length is equal

$$\frac{1.47}{1.99} 10^{-28} \, cm = 7.4 \times 10^{-29} \, cm.$$

For the electromagnetic field of the charged mass instead the radius of mass we need to take the radius of charge $r_c$ in the units of length,

$$r_c = \frac{q^2}{4 \pi \varepsilon_0 m_c c^2},$$

where $\varepsilon_0$ is electric constant, $q$ is elementary electric charge, $m_c$ is the mass of the charge.

For example, for the electromagnetic field in the hydrogen atom the radius of proton charge $r_c$ in the units of length is

$$r_c = \frac{q^2}{4\pi\varepsilon_0 m_p c^2},$$

where $m_p$ is the rest mass of proton. Using the radius of the atom $r_a$,

$$r_a = \frac{\varepsilon_0 \hbar^2}{\pi m_e q^2},$$

where $2\pi\hbar$ is Planck constant, $m_e$ is electron rest mass, we can obtain the relation $r_c/r_a$,

$$\frac{r_c}{r_a} \cong 2.9\times 10^{-8}.$$

For the comparison we find the relations $r_M/p$ and $r_M/a$ for the Mercury – the nearest planet to the Sun,

$$\frac{r_M}{p} \cong 3.2\times 10^{-8}, \quad \frac{r_M}{a} \cong 2.1\times 10^{-8}.$$

Here $p$ – the perihelion, $a$ – the aphelion of the Mercury. These data tell about some analogy between the electromagnetic and gravitational fields.

Let us find the relation between the radius of proton charge $r_{cp}$ to the radius of proton mass $r_{mp}$, where

$$r_{cp} \equiv \frac{q^2}{4\pi\varepsilon_0 m_p c^2}, \quad r_{mp} \equiv \frac{G m_p}{c^2}.$$

We get

$$\frac{r_{cp}}{r_{mp}} = \frac{q^2}{4\pi\varepsilon_0 G m_p^2} = 1.23587 * 10^{36}.$$

The obtained number is the relation between the electromagnetic and gravitational forces.

## 10. FRIEDMANN-LOBACHEVSKY MODEL.

### 10.1. The metric and the Ricci matrices.

The Friedmann-Lobachevsky model [4], is starting from the assumption that the expressions for the metric matrix is written in the form,

$$g = f^2(s) G,$$

where

$$s \equiv \sqrt{\tilde{x} G x},$$

$$\frac{ds}{d\tau} = \frac{\tilde{x} G u(x)}{s},$$

$$G = \begin{pmatrix} -1 & 0 & 0 & 0 \\ 0 & -1 & 0 & 0 \\ 0 & 0 & -1 & 0 \\ 0 & 0 & 0 & 1 \end{pmatrix}.$$

## 10.2 The Christoffel matrices of the second kind.

For this metric matrix we find the Christoffel matrices of the second kind,

$$\sigma^1 = \frac{f'(s)}{s f(s)} \begin{pmatrix} -x_1 & -x_2 & -x_3 & x_4 \\ x_2 & -x_1 & 0 & 0 \\ x_3 & 0 & -x_1 & 0 \\ x_4 & 0 & 0 & -x_1 \end{pmatrix},$$

$$\sigma^2 = \frac{f'(s)}{s f(s)} \begin{pmatrix} -x_2 & x_1 & 0 & 0 \\ -x_1 & -x_2 & -x_3 & x_4 \\ 0 & x_3 & -x_2 & 0 \\ 0 & x_4 & 0 & -x_2 \end{pmatrix},$$

$$\sigma^3 = \frac{f'(s)}{s f(s)} \begin{pmatrix} -x_3 & 0 & x_1 & 0 \\ 0 & -x_3 & x_2 & 0 \\ -x_1 & -x_2 & -x_3 & x_4 \\ 0 & 0 & x_4 & -x_3 \end{pmatrix},$$

$$\sigma^4 = \frac{f'(s)}{s f(s)} \begin{pmatrix} x_4 & 0 & 0 & -x_1 \\ 0 & x_4 & 0 & -x_2 \\ 0 & 0 & x_4 & -x_3 \\ -x_1 & -x_2 & -x_3 & x_4 \end{pmatrix}.$$

## 10.3 The equations of motion.

It is possible to verify that geodesic equation

$$\boxed{\frac{du(x)}{d\tau} = -\sigma^\alpha u(x) u_\alpha(x)},$$

for the metric matrix

$$g = f^2(s)G$$

is satisfied, if we take $u(x)$ in the form

$$u(x) = \frac{dx}{d\tau} = \frac{x}{s f(s)}.$$

For this $u(x)$ we obtain

$$\frac{ds}{d\tau} = \frac{\tilde{x} G u(x)}{s} = \frac{1}{f(s)}$$

$$\frac{d u(x)}{d\tau} = \frac{d}{d\tau}(\frac{x}{s f(s)}) = -\frac{x f'(s)}{s f^3(s)}.$$

$$\frac{d u(x)}{d\tau} = -\sigma^\alpha u(x) u_\alpha(x) = -\frac{x f'(s)}{s f^3(s)}$$

.
Using the equations

$$u(x) = \frac{dx}{d\tau} = \frac{x}{s f(s)}$$

$$\frac{ds}{d\tau} = \frac{1}{f(s)}.$$

we obtain

$$\frac{dx}{s} = \frac{x}{s}.$$

The solution of the last differential equation is

$$x_\lambda = a_\lambda s, \quad \alpha = 1, 2, 3,$$

where $a_\lambda$ is *constant* and

$$a_4 = \sqrt{1 + a_1^2 + a_2^2 + a_3^2}.$$

The dimensionless 3-velocities are introduced,

$$\beta_\alpha \equiv \frac{dx_\alpha}{dx_4} = \frac{u_\alpha(x)}{u_4(x)} = \frac{a_\alpha}{a_4}, \quad \alpha = 1, 2\ 3;.$$

$$\beta \equiv \frac{\sqrt{a_1^2 + a_2^2 + a_3^2}}{a_4} = \sqrt{\beta_1^2 + \beta_2^2 + \beta_3^2}.$$

Therefore we have

$$a_4 = \frac{1}{\sqrt{1 - \beta^2}}.$$

$$x_\lambda = \frac{\beta_\lambda}{\sqrt{1 - \beta^2}} s, \quad \lambda = 1, 2\ 3;$$

$$x_4 = \frac{s}{\sqrt{1 - \beta^2}}.$$

Since $s^2 > 0$ we must have $\beta^2 < 1$.

## 10.4 The expression for the Ricci matrix.

The matrix $g^{-1} R$, where $R$ is the Ricci matrix, has four eigenvalues,

$$\mu_1, \mu_2, \mu_3, \mu_4,$$

three of them are equal

:

$$\mu_1 = \mu_2 = \mu_3 = h_1, \quad \mu_4 = h_2,$$

where

$$\mu_1 \equiv \frac{s(f'(s))^2 + 5f(s) f'(s) + s f(s) f''(s)))}{s f^4(s)},$$

$$\mu_4 \equiv \frac{3(-s(f'(s))^2 + f(s)(f'(s) + s f''(s)))}{s f^4(s)}.$$

The expressions for the absolute and coordinate Ricci matrices have the form:

$$K = \mu_4 \, U \tilde{U} + \mu_1 \, (I - U \tilde{U}),$$

$$R = \frac{1}{(s f(s))^2} (\mu_4 - \mu_1) \, g \, x \tilde{x} \, g + \mu_1 \, g = (\mu_4 - \mu_1) \, g \, u(x) \tilde{u}(x) \, g + \mu_1 \, g,$$

where

$$u(x) = \frac{x}{s\,f(s)}.$$

### 10.5 The field matrix.

Using the expression,

$$u(x) = \frac{dx}{d\tau} = \frac{x}{s\,f(s)}$$

we find the expressions for the field matrix,

$$P = h(I - U\tilde{U}),$$

$$\underline{P} = h(g - g\,u(x)\,\tilde{u}(x)\,g),$$

where

$$h \equiv \frac{f(s) + s\,f'(s)}{s\,f^{2}(s)} = -s\,\frac{d}{ds}\left(\frac{1}{s\,f(s)}\right).$$

$$\operatorname{Tr} P = \langle P \rangle = \langle g^{-1}\,\underline{P}\rangle = 3h,$$

### 10.6 Solving the matrix metric equation.

We will consider the solution of the matrix metric equation for two different cases: the first case, when all eigenvalues of the matrix $g^{-1}R$ are equal, and the second case, when the coordinate velocity vector $u$ $(x)$ is equal to the eigenvector associated with the eigenvalue $\mu_4$ of the matrix $g^{-1}R$.

### 10.6.1 The first solution: the case of a maximally uniform space.

The first solution corresponds to all equal eigenvalues of the matrix $g^{-1}R$,

$$\mu_1 = \mu_2 = \mu_3 = \mu_4 = \rho.$$

For this case

$$\mu_1 - \mu_4 = 0,$$

$$g^{-1}R = \rho I,$$

$$g^{-1}R\, u(x) = \rho\, u(x).$$

where the coordinate velocity vector $u$ $(x)$ equals to the eigenvector associated with the eigenvalue $\mu_4$ of the matrix $g^{-1}R$,

$$R\, u(x) = \mu_4\, g\, u(x).$$

The equation

$$\mu_1 - \mu_4 = \frac{2(-2s(f'(s))^2 + f(s)(-f'(s) + s f''(s)))}{s f^4(s)} = 0$$

will be satisfied, if

$$f(s) = \frac{1}{a - b s^2},$$

$$\rho = \mu_1 = 12\,ab.$$

where $a$ and $b$ are constants and therefore the density $\rho$, cosmological constant, is also constant. Putting $a = 1$ we get

$$b = \frac{\rho}{12},$$

so that

$$f(s) = \frac{1}{1 - \frac{\rho}{12}s^2}.$$

The twice covariant Ricci matrix in this variant has the following form,

$$R = \rho\,g.$$

This is the case of a maximally uniform space [4]. In the considered case the constant of curvature $k$,

$$k = \frac{1}{12}\langle K \rangle = \frac{1}{12}\langle g^{-1} R \rangle = \frac{\rho}{3}.$$

Using the obtained results we can write

$$f(s) = \frac{1}{1 - \frac{k}{4}s^2} = \frac{1}{1 - \frac{\rho}{12}s^2},$$

$$g = \frac{1}{1-\frac{k}{4}s^2}G = \frac{1}{1-\frac{\rho}{12}s^2}G.$$

We have received the same result, as the similar result in [4], which was obtained using the Einstein equations of gravitation.

In the Friedmann-Lobachevsky space with the metric matrix g,

$$g = f^2(s)G,$$

$$f(s) = (1+\frac{a}{s})^2,$$

$$x_\lambda = \beta_\lambda x_4,$$

$$s = x_4\sqrt{1-\beta^2}.$$

$$a = s_m = x_{4m}\sqrt{1-\beta^2}.$$

each mass moves with a constant velocity proportional to its coordinates and therefore in this space all distances increase proportionately with time.

### 10.6.2 The second solution: the coordinate velocity vector $u(x)$ is equal to the eigenvector associated with the eigenvalue $\mu_4$ of the matrix $g^{-1}R$.

The matrix metric equation

$$g^{-1}R\,u(x) = \mu_4\,u(x),$$

is satisfied, because

$$R = (\mu_4 - \mu_1)\, g\, u(x)\, \tilde{u}(x)\, g + \mu_1\, g,$$

$$\tilde{u}(x)\, g\, u(x) = 1,$$

$$R\, u(x) = \mu_4\, g\, u(x),$$

$$g^{-1} R\, u(x) = \mu_4\, u(x).$$

Here

$$\mu_4 = \rho(s) = \frac{3(-s\,(f'(s))^2 + f(s)(f'(s) + s\, f''(s)))}{s\, f^4(s)}.$$

From the equation of continuity

$$\tilde{D}(\rho U) = \tilde{\partial}_{\downarrow}\, e^{-1}\,(e\,(\rho\, u(x))_{\uparrow}) = \tilde{i}\,(\lambda)\,(\sigma^\lambda\, \rho u(x) + \partial_\lambda\,(\rho u(x))),$$

where

$$\rho u(x) = \rho(s)\,\frac{x}{s\, f(s)},$$

we obtain

$$\tilde{D}(\rho(s) U) = \frac{3\rho(s) f'(s) + f(s)(\dfrac{3\rho(s)}{s} + \rho'(s))}{f^2(s)}$$

Solving the equation of continuity,

$$\tilde{D}(\rho(s)U) = \frac{3\rho(s)f'(s) + f(s)(\frac{3\rho(s)}{s} + \rho'(s))}{f^2(s)} = 0,$$

we find $\rho(s)$,

$$\rho(s) = \frac{c_1}{(s f(s))^3}.$$

Solving the equation

$$\rho(s) = \mu_4 = \frac{3(-s(f'(s))^2 + f(s)(f'(s) + s f''(s)))}{s f^4(s)}$$

we obtain the following expressions for the functions $f(s)$ and $\rho(s)$,

$$f(s) = \frac{3 c_1 (1 + (\frac{s}{s_m})^{d/3})^2}{2 d^2 s (\frac{s}{s_m})^{d/3}},$$

$$\rho(s) = \frac{8 d^6}{27 c_1^2} \frac{(\frac{s}{s_m})^d}{(1 + (\frac{s}{s_m})^{d/3})^6} = \rho_m \frac{64 (\frac{s}{s_m})^d}{(1 + (\frac{s}{s_m})^{d/3})^6}.$$

Here $\rho_m$, $s_m$, $c_1$ and $d$ are constants. The density $\rho(s)$ has maximum $\rho = \rho_m$ at $s = s_m$, where

$$s_m = x_{4m}\sqrt{1 - \beta^2}.$$

$$\rho(s) = \rho_m \frac{64(\frac{s}{s_m})^d}{(1+(\frac{s}{s_m})^{d/3})^6}$$

$$\rho_m = \frac{d^6}{216 c_1^2}.$$

Therefore, the expression for $f(s)$ will be

$$f(s) = \frac{d(1+(\frac{s}{s_m})^{d/3})^2}{4s\sqrt{6\rho_m}(\frac{s}{s_m})^{d/3}}.$$

The general form of the metric matrix at this case becomes

$$g = \frac{d^2(1+(\frac{s}{s_m})^{d/3})^4}{96 s^2 \rho_m (\frac{s}{s_m})^{2d/3}} G.$$

**10.7 The mathematical model of the Big Bang.**

Comparing with the measurements of the cosmic microwave background spectrum from COBE, it is possible to suppose that the expressions

$$g = \frac{d^2(1+(\frac{s}{s_m})^{d/3})^4}{96 s^2 \rho_m (\frac{s}{s_m})^{2d/3}} G,$$

$$\rho(s) = \rho_m \frac{64(\frac{s}{s_m})^d}{(1+(\frac{s}{s_m})^{d/3})^6},$$

present the mathematical model of the Big Bang.

Fig.1 The comparison of our mathematical model
with the cosmic microwave background spectrum from COBE.

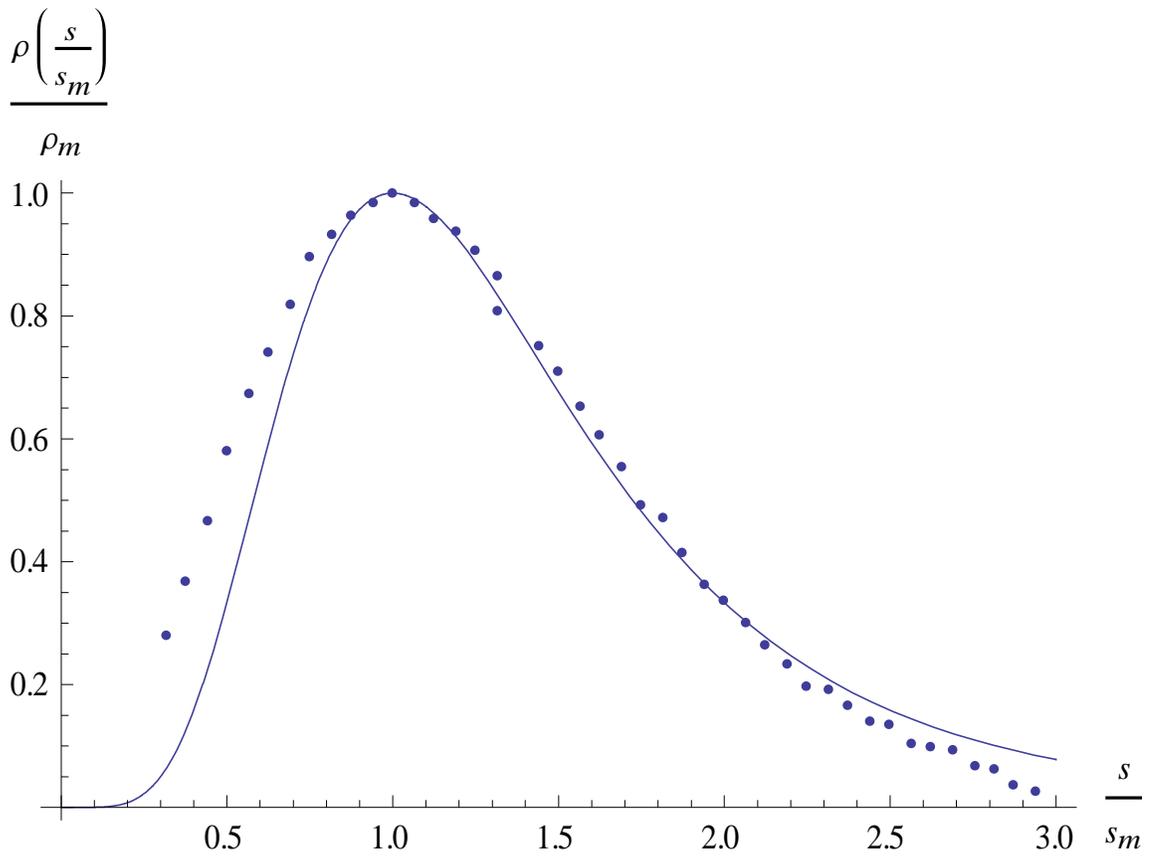

Fig.1 shows two curves. The continuous line is our function $\dfrac{\rho(\dfrac{s}{s_m})}{\rho_m}$ for $d=5.4$. The dotted line is the cosmic microwave background spectrum from COBE, where $\rho$ is the intensity and $s$ is wave lengths/centimetre [5].

## 11. THE FLAT N–DIMENSIONAL SPACE.

### 11.1 The definition of the flat n–dimensional space.

In the tensor classical theory of general relativity the space becomes flat if all elements of the Riemann curvature tensor $R^m_{kba}$, $m,k,b,a = 1,...,n$ become zero. In the matrix language the $n$ – dimensional space becomes flat if all elements of the two-index coordinate cotravariant–covariant *Riemann curvature matrix* $\sigma^{ab}$ become zero.

### 11.2 The coordinate frame matrix $e_0$ for the flat n-dimensional sapace.

Is it possible to find such coordinate frame matrix

$$e_0 = e_{0\,flat}$$

as a function of all elements of the coordinate $n$-vector $x$, for which the *Riemann curvature matrix* $\sigma^{ab}$ is zero? Yes, and this matrix is

$$e_{0\,flat} \equiv I + F_a(x_k)(I - I(k))\, x\, \tilde{i}(k)$$

where $F_a(x_k)$ is an antisymmetric $n\times n$ matrix function of $x_k$. For this coordinate frame matrix we find

$$\sigma^k_{flat} = e^{-1}_{0\,flat}(F_a(x_k)\,e_{0\,flat} + \partial_k\,e_{0\,flat})$$

$$\sigma^m_{flat} = e^{-1}_{0\,flat} \, \partial_m \, e_{0\,flat}), \quad m \neq k$$

Taking into the consideration the equation for the orthogonal matrix $\Omega$,

$$\tilde{\Omega} \partial_a \Omega = (e_0 \, \sigma^a - \partial_a \, e_0) e_0^{-1}, \quad a = 1,...,n$$

for the flat space we find

$$\tilde{\Omega}_{flat} \, \partial_m \, \Omega_{flat} = 0, \quad m \neq k$$

$$\tilde{\Omega}_{flat} \, \partial_k \, \Omega_{flat} = F_a(x_k), \quad k = 1,...,n$$

Therefore the orthogonal matrix function $\Omega_{flat}$ depends only on the variable $x_k$ and is determined by the differential equation

$$\frac{d\Omega_{flat}}{dx_k} = \Omega_{flat} \, F_a(x_k)$$

### 11.3 The absolute frame matrix *e* for the flat *n*–dimensional space.

The absolute frame matrix *e* for the flat *n*–dimensional space takes the form:

$$e_{flat} = \Omega_{flat} \, e_{0\,flat}$$

or

$$e_{flat} = \Omega_{flat} \, (I + F_a(x_k)(I - I(k)) \, \tilde{xi}(k))$$

### 11.4 The coordinate metric matrix *g* for the flat *n*–dimensional space.

The coordinate metric matrix $g$ for the flat $n$–dimensional space is denoted by $g_{flat}$, where

$$g_{flat} \equiv \tilde{e}_{0\,flat} e_{0\,flat}$$

## 11.5 The connection between the full differentials $dQ$ and $dx$ for the flat $n$–dimensional space.

The connection between the full differentials $dQ$ and $dx$

$$dQ = e\, dx$$

for the flat $n$–dimensional space becomes

$$dQ_{flat} = \Omega_{flat}\ (I + F_a(x_k)(I - I(k))\, x\, \tilde{i}(k))\, dx$$

Therefore the quadratic form

$$\boxed{d\tau^2 = d\tilde{x}\ g_{flat}\ dx}$$

is reducible to

$$\boxed{d\tau^2 = d\tilde{Q}_{flat}\ dQ_{flat}}$$

Here $Q$ can be presented as the sum of two vector functions,

$$Q_{flat} = M + W,$$

where

$$M \equiv \int_0^{x_k} \Omega_{flat}\, dx_k$$

$$W \equiv \Omega_{flat}(I - I(k))x,$$

$$dM = \Omega_{flat} I(k) dx,$$

$$dW = d\Omega_{flat}(I - I(k))x + \Omega_{flat}(I - I(k))dx =$$
$$= \Omega_{flat} F_a(x_k)(I - I(k))x\tilde{i}(k)dx + \Omega_{flat}(I - I(k))dx.$$

From the last two equations it follows the expression for $dQ_{flat}$,

$$dQ_{flat} = dM + dW = \Omega_{flat}(I + F_a(x_k)(I - I(k))x\tilde{i}(k))dx,$$

which proves that the solution of the differential equation

$$dQ_{flat} = \Omega_{flat}(I + F_a(x_k)(I - I(k))x\tilde{i}(k))dx$$

is

$$Q_{flat} = M + W = \int_0^{x_k} \Omega_{flat} dx_k + \Omega_{flat}(I - I(k))x$$

where the orthogonal matrix function is defined by the differential equation

$$\frac{d\Omega_{flat}}{dx_k} = \Omega_{flat} F_a(x_k).$$

## 12. SUMMARY

Instead the tensor theory a new matrix approach to the theory of field in *n*-dimensional metric space is developed. Two spaces are considered: the *n*-dimensional Riemannian space and the *n*-dimensional space of non-integrable vectors and matrices. We call the last space as the *absolute* space. The *absolute* (invariant) field matrix is introduced and two forms of the matrix field equations are derived.

Applications to the four-dimensional Riemannian spacetime are considered. It is shown that in this case the absolute field matrix is the absolute electromagnetic-gravitational field matrix. The symmetric part of this matrix is the absolute gravitational field matrix. The antisymmetric part is the absolute electromagnetic field matrix. The differential equations for these two field matrices and for the metric matrix have been obtained. The partial case of these equations presents the Maxwell equations for the electromagnetic field.

Six general solutions for the static spherically symmetric gravitational field are found. The first solution is the most general and it contains three constants, which can be interpreted as two masses and density. It describes the motion of the mass *m* in the field of the mass *M* and the density $\rho$. The second solution describes the motion of the mass *m* in the field of the mass *M*. The third solution describes the motion of the mass *m* in the field of the density $\rho$. The fourth solution describes the motion of the very small mass *m*<<*M* (*m*=0) in the field of the mass *M* and the density $\rho$. The fifth solution describes the motion of the very small mass *m*<<*M* (*m*=0) in the field of the mass *M*. The sixth solution is the general form of spherically symmetric solution for $m = 0$ and $\rho = 0$ with the metric matrix $g(x_3)$, where the matrix element $g_{33}(x_3)$ is an arbitrary function.

The Schwarzschild diagonal metric matrix $g_s$ is the particular case of the sixth solution, where the function $g_{33}(x_3)$ is taken in the form:

$$g_{33}(x_3) = -\frac{1}{1 - \frac{2r_M}{x_3}},$$

and which has a singularity at the Schwarzschild radius $x_3 = 2r_M$.

Our sixth solution has no an apparent singularity at the Schwarzschild radius. For this solution the formulae for the minimum and maximum orbital velocity and for the perihelion precession of planets through perihelion and aphelion distances are given. In our calculations of extreme velocities of planets all first four digits coincide with the NASA data.

The obtained matrix field equations were solved also for the Friedmann-Lobachevsky model of the metric *g*. We obtained the expression for the density of particles which looks like the expression describing the mathematical model of the Big Bang, where the density at the beginning of time is zero and after that it is growing very fast until very big maximum and after this maximum it is slowly decreasing. Our expression is compared with the experimental cosmic microwave background spectrum from COBE.

## REFERENCES

[1] L.D. Landau and E.M. Lifshitz, The Classical Theory of Fieds, Butterworth-Heinemann Ltd. Oxford, London, 1994

[2] A. Einstein, The Meaning of Relativity. Princeton University Press, Princeton, New Jersey, 1953

[3] S. Hawking and R. Penrose, The Nature of Space and Time. Princeton University Press, Princeton, New Jersey, 1996.

[4] V.A. Fok, The Theory of Space, Time and Gravitation. Macmillan Co./Pergamon Press, New York, 1964

[5] D.J. Finsen, E.S.Cheng, J.M.Gales, J.C.Mather, R.A. Shafer, E.L. Wright, The Cosmic Microwave Background Spectrum from the Full COBE/FIRAS Data Set, Astrophys.J. 473 (1996) 576